\documentclass[preprint,showpacs,preprintnumbers,amsmath,amssymb,floatfix]{revtex4}

\usepackage{bm}% bold math

\usepackage{graphicx}
\usepackage{dcolumn}

\begin{document}
%
%\setcounter{page}{1}

%%%%draft

\title{
The Tenth-Order QED Contribution to the Lepton g-2: \\
Evaluation of Dominant $\alpha^5$ Terms of Muon g-2
}

\author{Toichiro Kinoshita }
\email{tk@hepth.cornell.edu}
\affiliation{Laboratory for Elementary-Particle Physics\\ Cornell
University,
         Ithaca, New York, 14853  }

\author{Makiko  Nio}
\email{nio@riken.jp}
\affiliation{Theoretical Physics Laboratory,
RIKEN, Wako, Saitama, Japan 351-0198 }

\date{\today}
\begin{abstract}
The QED contribution to the anomalous magnetic moments of electron
and muon are known very precisely up to the order $\alpha^4$.
However, the knowledge of $\alpha^5$ term
will also be required when the precision of measurement improves further.
This paper reports the first systematic attempt to evaluate
the $\alpha^5$ term.
Feynman diagrams contributing to this term can be classified 
into six gauge-invariant sets which can be subdivided further
into 32 gauge-invariant subsets.
Thus far we have numerically evaluated all integrals of 17 
gauge-invariant subsets which contain light-by-light-scattering
subdiagrams and/or vacuum-polarization subdiagrams.
They cover most of leading terms of muon $g-2$
and lead to a preliminary result
663 (20) $(\alpha/\pi )^5$, which is 8.5 times more
precise than the old estimate.

\end{abstract}

\maketitle

%%\narrowtext

%%%%%%%%%%%%%%%%%%%%%%%%%%%%%%%%%%%%%%%%%%%%%%%%%%%%%%%%%%
\section{Introduction}
\label{sec:intro}
%%%%%%%%%%%%%%%%%%%%%%%%%%%%%%%%%%%%%%%%%%%%%%%%%%%%%%%%%%

The deviation of the electron $g$ value from 2 predicted
by Dirac's theory was first confirmed by an experiment
on atomic spectrum \cite{kusch}.
Schwinger showed that this deviation can be explained
as the effect of radiative correction by the relativistic
renormalized QED which he had developed \cite{schwinger}.
Together with the discovery of Lamb shift in the spectrum
of hydrogen atom \cite{lamb}, it provided convincing experimental
evidence that (until then discredited) QED is 
capable of predicting the effect
of electromagnetic interaction precisely, provided that it is
renormalized.

%%%%%%%%%%%%%%%%%%%%%%%%%%%%%%%%%%%%%%%%%%%%%%%%%%%%%%%%%%
\subsection{Measurement of electron g-2}
\label{sec:measeg-2}
%%%%%%%%%%%%%%%%%%%%%%%%%%%%%%%%%%%%%%%%%%%%%%%%%%%%%%%%%%

By 1970's the precision of measurement of electron g-2
was improved by four more orders of magnitude by means of spin precession
of the electron moving in a constant uniform magnetic field \cite{rich}.
The value of the electron g-2 was improved further by
three additional orders of magnitude in a Penning trap experiment
by Dehmelt's group at the University of Washington.
Their published results are \cite{vandyck}
\begin{eqnarray}
a_{e^-} &=& 1~159~652~188.4~(4.3) \times 10^{-12}~~~[3.8~ppb], \nonumber  \\
a_{e^+} &=& 1~159~652~187.9~(4.3) \times 10^{-12}~~~[3.8~ppb],
\label{ae-seattle}
\end{eqnarray}
where the numerals 4.3 in parentheses represent
the combined statistical and systematic 
uncertainties in the last digits of the measured value.
$1~{\rm ppb} = 10^{-9}$.

The precision of measurement has thus been improved by seven orders of
magnitude over 40 years.
This enormous improvement in measurement 
was matched by the improvement of theory of radiative
correction to the electron g-2 from the order
$\alpha$ to the  order $\alpha^4$, leading to the most
stringent test of the validity of QED.

The uncertainty of the experiment (\ref{ae-seattle}) was dominated
by the cavity shift due to the interaction of the electron
with the hyperboloid cavity, which has a very complicated
resonance structure.
Several efforts were made to reduced this uncertainty \cite{vandyck2,mittleman}.
One of them is to replace the hyperboloid cavity by a
cylindrical cavity, which allows analytic computation of
the structure of the resonance \cite{lbrown}.
Gabrielse's new measurement of the electron g-2 is based
on this analysis.
Recently a preliminary result of this measurement was reported, %:\cite{odom}
%
%\begin{equation}
%a_{e^-} = 1~159~652~180.86~(0.57) \times 10^{-12}, 
%\label{ae-harvard}
%\end{equation}
%
which is 7.5 times more precise than (\ref{ae-seattle})\cite{odom}.

%%%%%%%%%%%%%%%%%%%%%%%%%%%%%%%%%%%%%%%%%%%%%%%%%%%%%%%%%%
\subsection{Theory of electron g-2 to order $\alpha^4$}
\label{sec:theg-2}
%%%%%%%%%%%%%%%%%%%%%%%%%%%%%%%%%%%%%%%%%%%%%%%%%%%%%%%%%%

The QED contribution to the electron g-2 can be written as
\begin{equation}
a_e (QED) =A_1 +
A_2 (m_e /m_\mu )+
A_2 (m_e /m_\tau )+
A_3 (m_e /m_\mu, m_e /m_\tau )
\label{aeQED}
\end{equation}
and $A_i,~i=1,2,3$, can be expanded as
\begin{equation}
A_i = A_i^{(2)} \left(\frac{\alpha}{\pi}\right) +
A_i^{(4)} \left(\frac{\alpha}{\pi}\right)^2 +
A_i^{(6)} \left(\frac{\alpha}{\pi}\right)^3 + \ldots.
\end{equation}
The first four coefficients of $A_1$ are 
\begin{eqnarray}
A_1^{(2)} &=& 0.5,
\nonumber  \\
A_1^{(4)} &=& -0.328~478~965 \ldots ,
\nonumber  \\
A_1^{(6)} &=& 1.181~241~456 \ldots ,
\nonumber  \\
A_1^{(8)} &=& -1.728~3~(35) .
\end{eqnarray}
$A_1^{(2)}$ and $A_1^{(4)}$ are known analytically \cite{schwinger,petermann,sommerfield}.
$A_1^{(6)}$ was obtained by both numerical \cite{kino1} and analytic integrations \cite{laporta1}.
$A_1^{(8)}$ is obtained thus far by numerical integration only \cite{kn5}.
Its uncertainty has been reduced by 10 compared with the old one \cite{hugheskinoshita}.
Although it has been evaluated by one method only,
it has been subjected to an extensive cross-checking among diagrams of 8th-order
and also with 6th-, 4th-, and 2nd-order diagrams.

$A_2$ terms are small:
\begin{eqnarray}
A_2^{(4)}(m_e/m_\mu ) (\alpha/\pi )^2 &=& 2.804 \times 10^{-12},
\nonumber  \\
A_2^{(4)}(m_e/m_\tau ) (\alpha/\pi )^2 &=& 0.010 \times 10^{-12},
\nonumber  \\
A_2^{(6)}(m_e/m_\mu ) (\alpha/\pi )^3 &=& -0.924 \times 10^{-13},
\nonumber  \\
A_2^{(6)}(m_e/m_\tau ) (\alpha/\pi )^3 &=& -0.825 \times 10^{-15}
\end{eqnarray}
The contribution of $A_3$ term is even smaller ($\sim 2.4 \times 10^{-21}$).
The non-QED contribution of the Standard Model are also known \cite{jeger,krause,czarnecki}
\begin{eqnarray}
a_e ({\rm hadron}) &=& 1.671~(19) \times 10^{-12},
\nonumber  \\
a_e ({\rm weak}) &=& 0.030~(1) \times 10^{-12}.
\end{eqnarray}

To compare the theory with the measured value of $a_e$ one needs
a value of $\alpha$ obtained by some non-QED measurement.
The best available $\alpha$ at present is \cite{wicht,mohr}
\begin{equation}
\alpha^{-1} (h/M_{C_s}) = 137.036~000~1~(11)~~~~[7.4~ppb]~.
\label{cesium}
\end{equation}
This leads to 
\begin{eqnarray}
a_e (h/M_{C_s}) &=& 1~159~652~175.86~(0.10)(0.26)(8.48) \times 10^{-12},
\nonumber  \\
a_e (exp) &-& a_e (h/M_{C_s}) = 12.4~(9.5) \times 10^{-12},
\end{eqnarray}
where 0.10 is the remaining uncertainty of the $\alpha^4$ term,
0.26 is based on an educated guess ($A_1^{(10)}=0(3.8)$)
made by Mohr and Taylor \cite{mohr},
and 8.48 is the uncertainty in the measurement (\ref{cesium}).
The error 8.48 is still large but is within a factor 2 of the error
of the Seattle measurement (\ref{ae-seattle}).

%%%%%%%%%%%%%%%%%%%%%%%%%%%%%%%%%%%%%%%%%%%%%%%%%%%%%%%%%%
\subsection{Measurement of muon g-2}
\label{sec:measmug-2}
%%%%%%%%%%%%%%%%%%%%%%%%%%%%%%%%%%%%%%%%%%%%%%%%%%%%%%%%%%

The last and best of three measurements of the muon g-2 at CERN had an uncertainty
of 7 ppm \cite{farley}.
After years of hard work the muon g-2 measurement at the
Brookhaven National Laboratory has come close to the
design goal (0.35 ppm):
\cite{bennett2}
\begin{equation}
 a_{\mu^-}({\rm exp}) 
  = 11~659~214~(8)~(3) \times 10^{-10} ~~~~~~~~(0.7 {\rm ~ppm}).
\label{amuexp00}
\end{equation}
The world average of $a_\mu (\rm{exp})$ obtained
by combining this and earlier measurements \cite{farley,bennett1,brown1,brown2} is 
\begin{equation}
 a_{\mu}({\rm exp})
  = 11~659~208~(6) \times 10^{-10} ~~~~~~~~(0.5 {\rm ~ppm}).
\label{amuexpall}
\end{equation}
%
%This provides the most stringent test of the
%Standard Model. 
%
%[Next two paragraphs must be rewritten.]

%%%%%%%%%%%%%%%%%%%%%%%%%%%%%%%%%%%%%%%%%%%%%%%%%%%%%%%%%%
\subsection{Hadronic and electroweak contributions to muon g-2}
\label{sec:thmug-2}
%%%%%%%%%%%%%%%%%%%%%%%%%%%%%%%%%%%%%%%%%%%%%%%%%%%%%%%%%%

Currently, the prediction of the Standard Model 
reflects the difficulty in the treatment of the hadronic contribution
\cite{czarnecki1,davier1,narison1,troconiz1,knecht2,ramsey}.
The lowest-order hadronic vacuum-polarization 
effect on $a_\mu$  has thus far been determined from three sources,

(i) $e^+ e^-$ annihilation cross section, 

(ii) hadronic $\tau$ decays,

(iii) $e^+ e^-  \rightarrow \gamma + {\rm hadrons}$ ~~~  [radiative return].

The process (i) has been analyzed by many groups over years.
Some recent results are \cite{hoecker,hagiwara,troconiz}
\begin{eqnarray}
 a_{\mu}({\rm had.LO})
  &=& 6934~(53)_{exp}~(36)_{rad} \times 10^{-11}, ~\nonumber  \\
 a_{\mu}({\rm had.LO})
  &=& 6924~(59)_{exp}~(24)_{rad} ~\times 10^{-11},  \nonumber  \\ 
 a_{\mu}({\rm had.LO})
  &=& 6944~(48)_{exp}~(10)_{rad} ~\times 10^{-11}. 
\label{amuhadLO}
\end{eqnarray}

\noindent
Recent estimate of hadronic light-by-light scattering contribution is
\cite{melnikov}
\begin{equation}
a_{\mu}({\rm had.NL}) = 136~(25) \times 10^{-11} .
\label{hadNL}
\end{equation}

\noindent
The electroweak interaction effect is known to two-loop order:
\cite{knecht,czarnecki2}
\begin{eqnarray}
a_{\mu}(weak) &=& 152~(1) \times 10^{-11}, \nonumber   \\
a_{\mu}(weak) &=& 154~(1)~(2) \times 10^{-11},
\label{newweak1}
\end{eqnarray}
where (1) and (2) in the second line are estimates of
remaining theoretical uncertainty and
Higgs mass uncertainty, respectively.

%%%%%%%%%%%%%%%%%%%%%%%%%%%%%%%%%%%%%%%%%%%%%%%%%%%%%%%%%%
\subsection{QED contribution to muon g-2 to order $\alpha^4$}
\label{sec:mug-2QED}
%%%%%%%%%%%%%%%%%%%%%%%%%%%%%%%%%%%%%%%%%%%%%%%%%%%%%%%%%%

The QED contribution to the muon g-2 can be written as
\begin{equation}
a_\mu (QED) =A_1 +
A_2 (m_\mu/m_e )+
A_2 (m_\tau/m_e )+
A_3 (m_\mu/m_e , m_\tau/m_e ).
\label{amuQED}
\end{equation}
Renormalizability of QED guarantees that the functions $A_1$, $A_2$, and
$A_3$ can be expanded in power series in $\alpha / \pi $ with finite
calculable coefficients:
\begin{equation}
A_i~=~A_i^{(2)} \left( \frac{\alpha}{\pi} \right)~+~A_i^{(4)} \left(
 \frac{\alpha}{\pi} \right )^2 ~+~A_i^{(6)} \left( \frac{\alpha}{\pi}
\right)^3 ~+~.~.~.~,~~~~i~=~1,~2,~3.     \label{power}
\end{equation}
$A_1$ in (\ref{amuQED}) 
is identical with $A_1$ of (\ref{aeQED}) and 
has been evaluated to the eighth (i.e., ${\alpha}^4$) order.
As for $A_2$ and $A_3$,  it is
easy to see that $A_2^{(2)} = A_3^{(2)} = A_3^{(4)}=0$ since they have
no corresponding Feynman diagram.
$A_2^{(4)}$, $A_2^{(6)}$, and $A_3^{(6)}$ terms have been evaluated
accurately by power series expansion in $m_e /m_\mu$ or $m_e /m_\tau$.
Thus far $A_2^{(8)}$ and $A_3^{(8)}$ are known mostly
by numerical integration.

The QED contribution $ a_\mu ({\rm QED})$,
even though it is the predominant term of $a_\mu$,
has received little attention for many years
 because of its small error bars.
The theoretical uncertainty came predominantly from the
$\alpha^4$ term whose contribution to $a_\mu$ is about 3.3 ppm.
Recently we have completed a new evaluation of the $\alpha^4$ term
in which all contributing terms have been evaluated by two or more
independent calculations, 
uncovering an error in some diagrams \cite{kn2} and
eliminating the $\alpha^4$ term
as a possible source of theoretical uncertainty \cite{kn4,kn2}.
This causes the QED contribution shifted to 
%The new value of the QED contribution is
%
\begin{equation}
a_{\mu}(QED) = 116~584~719.43~(0.02)(1.15)(0.85) \times 10^{-11} ,
\label{newQEDvalue}
\end{equation}
where 0.03 is the remaining uncertainty of
the $\alpha^4$ term, 
an improvement of factor 40 over the previous result \cite{hugheskinoshita}.
The uncertainty 1.15 comes from a crude estimate of the contribution of
the $\alpha^5$ term, which now stands out as the largest source of uncertainty 
in $a_\mu$(QED).
The error 0.85 comes from the uncertainty in 
the value of $\alpha$ given in (\ref{cesium}).

The prediction of the Standard Model,
including the hadronic vacuum-polarization, hadronic light-by-light
contributions, and the electroweak effect, is
\begin{equation}
a_{\mu}(SM) = 116~591~870.7~(76.2) \times 10^{-11} ,
\label{amuSMnew}
\end{equation}
where the uncertainty in theory is mostly due to the hadronic
vacuum-polarization term.

%%%%%%%%%%%%%%%%%%%%%%%%%%%%%%%%%%%%%%%%%%%%%%%%%%%%%%%%%%
\subsection{Tenth-order term:  Why is it needed ?}
\label{sec:tenth}
%%%%%%%%%%%%%%%%%%%%%%%%%%%%%%%%%%%%%%%%%%%%%%%%%%%%%%%%%%

A very important byproduct of the study of $a_e$ is that it gives
the best value of the fine structure constant $\alpha$
available at present.
%%%If we use Odom's result we find
%%%%
%%%\begin{equation}
%%%\alpha^{-1} (a_e) = 137.035~999~708~(12)~(31)~(68)~~~[0.55~ppb]
%%%\label{odom}
%%%\end{equation}
%%%%
%%%where 12, 31, and 68 are the uncertainties of the $\alpha^4$ term,
%%%%$\alpha^5$ term, and experiment, respectively.
If we use the new experiment by the Harvard group, the precision of 
$\alpha$ is  almost an order of magnitude better than any
other measurement of $\alpha$.

Furthermore, the uncertainty of this measurement is only a factor 2 larger
than that of theory, which is mostly from the $\alpha^5$ term,
since the $\alpha^4$ term is known with small error.
Thus, when the measurement of $a_e$ is improved further,
reduction of the uncertainty of $\alpha^5$ term will become
crucial in order to obtain a better $\alpha (a_e)$.

For the muon the old estimate of $A_2^{(10)} (m_\mu/m_e )$
was 930 (170), which contributes only 0.054 ppm to $a_\mu$,
well within the current experimental uncertainty.
Thus improving $A_2^{(10)} (m_\mu/m_e )$ is not urgent.
However, it will become an important source of error in the
next generation of $a_\mu$ experiment.
This is why it is desirable to obtain a better value of
$A_2^{(10)} (m_\mu/m_e )$.  The preliminary value of $A_2^{(10)}$ was reported in
\cite{tau04}.

%%%%%%%%%%%%%%%%%%%%%%%%%%%%%%%%%%%%%%%%%%%%%%%%%%%%%%%%
\section{Classification of Tenth-Order Diagrams}
\label{sec:classification}
%%%%%%%%%%%%%%%%%%%%%%%%%%%%%%%%%%%%%%%%%%%%%%%%%%%%%%%%%%

Thus far only a small portion of tenth-order diagrams contributing to
the muon g-2 have been evaluated analytically \cite{laportaX},
or numerically \cite{oldtk1}.
Rough estimates based on the renormalization group and other considerations
have been made in order to identify leading terms \cite{oldtk1,karsh,yelli,
kataev}.

Of course an enormous amount of work is required
to go beyond this and evaluate $\alpha^5$ terms completely. 
Fortunately, for the muon g-2 the leading contribution comes from
those Feynman diagrams which contain $\ln (m_\mu/m_e)$ terms
whose sources can be readily identified as 
light-by-light-scattering subdiagrams and
vacuum-polarization insertions.
Thus relatively modest amount of work
will enable us to improve the value of $A_2^{(10)} (m_\mu/m_e) $
over the previous crude estimate.
On the other hand, the electron g-2, in particular $A_1^{(10)}$ term,
is much harder to evaluate.
Besides its gigantic size none of 12672 diagrams is dominant 
so that every term must be evaluated accurately.

For both $a_e$ and $a_\mu$ 
the first step is to count and classify Feynman diagrams
contributing to the $\alpha^5$ term. 
The contribution to the mass-independent term $A_1^{(10)}$
may be classified into six gauge-invariant sets, 
which further subdivided into 32 gauge-invariant subsets, depending on
the nature of subdiagrams (of the vacuum-polarization ({\it v-p}) type
or light-by-light-scattering ({\it l-l}) type).
Classification for $A_2^{(10)} (m_\mu/m_e)$ follows readily 
from that of $A_1^{(10)}$.  With help of the Feynman Diagram auto-generator
of GRACE system \cite{grace}  we count the number of diagrams belonging to each set.

%%%%%%%%%%%%%%%%%%%%%%%%%%%%%%%%%%%%%%%%%%%%%%%%%%%%%%%%%%
%\section{Clasification}
%\label{sec:clasification}
%%%%%%%%%%%%%%%%%%%%%%%%%%%%%%%%%%%%%%%%%%%%%%%%%%%%%%%%%%

%%%%%%%%%%%%%%%%%%%%%%%%%%%%%%%%%%%%%%%%%%%%%%%%%%%%%%%%%%
\subsection{Notation}
\label{sec:notation}
%%%%%%%%%%%%%%%%%%%%%%%%%%%%%%%%%%%%%%%%%%%%%%%%%%%%%%%%%%

Vacuum-polarization functions $\Pi$ needed for the evaluation of the
$\alpha^5$ contribution of the lepton $g-2$ may
be classified as follows:
In the following $\Pi_{x(y)}$ denotes a $\Pi_{x}$ containing
$\Pi_{y}$ on its internal photon line.

\vspace{2mm}
\noindent
$\Pi_{2}$, which consists of one closed lepton loop of second-order.

\vspace{2mm}
\noindent
$\Pi_{4}$, which consists of three proper
lepton loops of fourth order.

\vspace{2mm}
\noindent
$\Pi_{4(2)}$,
which consists of three diagrams of type $\Pi_4$ in which
one $\Pi_2$ is inserted in the internal photon line.

\vspace{2mm}
\noindent
$\Pi_{6}$,
which consists of 15 proper sixth-order lepton loops
which do not contain v-p loop.

\vspace{2mm}
\noindent
$\Pi_{6(2)}$,
which consists of 30 diagrams of type $\Pi_6$ in which
one $\Pi_2$ is inserted in the internal photon line.

\vspace{2mm}
\noindent
$\Pi_{4(4)}$,
which consists of 9 diagrams of type $\Pi_4$ whose
internal photon line contains a $\Pi_4$.

\vspace{2mm}
\noindent
$\Pi_{4(2,2)}$,
which consists of 3 diagrams of type $\Pi_4$ whose
internal photon line contains two $\Pi_2$'s.

\vspace{2mm}
\noindent
$\Pi_{8}$,
which consists of 105 proper eighth-order lepton loops.

\vspace{0.6cm}
Light-by-light scattering type functions $\Lambda$
needed for the evaluation of the
$\alpha^5$ contribution of the lepton $g-2$ may
be classified as follows:

\vspace{2mm}
\noindent
$\Lambda_{4}$,
which consists of six proper fourth-order lepton loop.

\vspace{2mm}
\noindent
$\Lambda_{4}^{(2)}$,
which consists of 60 diagrams in which lepton lines and vertices of
$\Lambda_4$ are modified by 2nd-order radiative corrections.

\vspace{2mm}
\noindent
$\Lambda_{4}^{(4)}$,
which consists of 105 diagrams in which lepton lines and vertices of
$\Lambda_4$ are modified by 4th-order radiative corrections.

\vspace{2mm}
\noindent
$\Lambda_{4(2)}^{(2)}$,
which consists of 60 diagrams in which
a v-p diagram $\Pi_2$ is inserted in the photon line
of $\Lambda_{4}^{(2)}$.

\vspace{2mm}
\noindent
$\Lambda_{6}$,
which consists of 120 proper lepton loops to which 6 photon lines
are attached.

\vspace{0.6cm}
Finally we need magnetic moment contributions of various orders.

\vspace{2mm}
\noindent
$M_2$, which consists of a lepton vertex diagram of second-order.

\vspace{2mm}
\noindent
$M_4$, which consists of six proper lepton vertices of fourth-order.

\vspace{2mm}
\noindent
$M_6$, which consists of 50 proper lepton vertices of sixth-order.

\vspace{2mm}
\noindent
$M_{6LL}$, which consists of 6 sixth-order vertex diagrams containing
an {\it external} light-by-light loop $\Lambda_4$.

\vspace{2mm}
\noindent
$M_8$, which consists of 518 proper lepton vertices of eighth-order
which have no closed lepton loops.

\vspace{2mm}
\noindent
$M_{8LLb}$,
obtained by replacing $\Lambda_4$ by $\Lambda_{4}^{(2)}$ in $M_{6LL}$,
and consists of 60 eighth-order vertex diagrams.

\vspace{2mm}
\noindent
$M_{8LLc}$,
obtained by attaching a virtual photon line to the open muon line
in all possible ways,
and consists of 48 eighth-order vertex diagrams.

\vspace{2mm}
\noindent
$M_{8LLd}$, obtained by inserting a $\Lambda_4$ {\it internally}
in the fourth-order $M_4$ in all possible ways.

%%%%%%%%%%%%%%%%%%%%%%%%%%%%%%%%%%%%%%%%%%%%%%%%%%%%%%%%
\subsection{Set I}
\label{sec:set1}
%%%%%%%%%%%%%%%%%%%%%%%%%%%%%%%%%%%%%%%%%%%%%%%%%%%%%%%%%%

\begin{figure}
\includegraphics[scale=0.8]{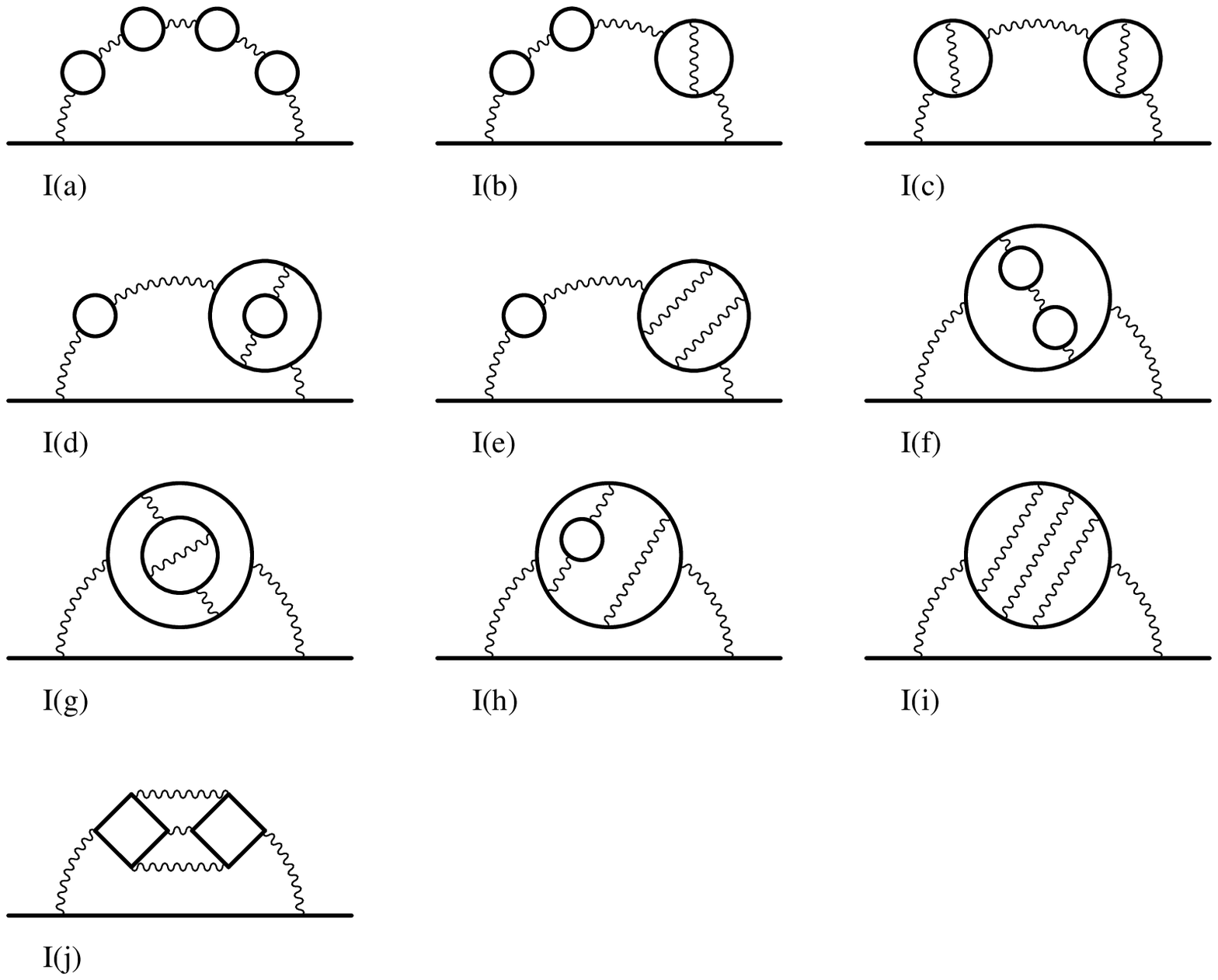}
\caption{\label{GroupI}  Set I}
\end{figure}

Diagrams of Set I are built from
the magnetic moment contribution $M_2$
of the second-order proper vertex.
It consists of 208 Feynman diagrams,
which can be classified further into
ten gauge-invariant subsets as indicated in Fig. \ref{GroupI}.
Let us mention only contributions to $A_1^{(10)}$ and
$A_2^{(10)}$ in this and subsequent sections
although many subsets contribute also to $A_3^{(10)}$.
This is because the contribution of $A_3^{(10)}$ is negligible
compared with others.

\vspace{2mm}
\noindent
{\em Subset I(a)}.
Diagrams obtained by inserting four $\Pi_2$'s
in $M_2$. One Feynman diagram belonging to this
subset contributes to $A_1^{(10)}$.
The contribution to $A_2^{(10)}$ comes from 15 diagrams ($2^4 -1 = 15$).

\vspace{2mm}
\noindent
{\em Subset I(b)}.
Diagrams obtained by inserting two $\Pi_2$'s and one $\Pi_4$
in $M_2$.
Nine Feynman diagrams of this subset contribute to $A_1^{(10)}$.
The contribution to $A_2^{(10)}$ comes from 63 diagrams ($9 (2^3 -1) =
63$).

\vspace{2mm}
\noindent
{\em Subset I(c)}.
Diagrams containing two $\Pi_4$'s
in $M_2$.
There are nine Feynman diagrams that contribute to $A_1^{(10)}$.
The contribution to $A_2^{(10)}$ comes from 27 diagrams ($9 (2^2 -1) =
27$).

\vspace{2mm}
\noindent
{\em Subset I(d)}.
Diagrams obtained by insertion of
one $\Pi_2$ and one $\Pi_{4(2)}$
in $M_2$.  Six Feynman diagrams contribute to $A_1^{(10)}$.
The contribution to $A_2^{(10)}$ comes from 42 diagrams ($6 (2^3 -1) =
42$).

\vspace{2mm}
\noindent
{\em Subset I(e)}.
Diagrams obtained by insertion of
one $\Pi_2$ and one $\Pi_6$
in $M_2$.  Thirty Feynman diagrams contribute to $A_1^{(10)}$.
The contribution to $A_2^{(10)}$ comes from 90 diagrams ($30 (2^2 -1) =
90$).

\vspace{2mm}
\noindent
{\em Subset I(f)}.
Diagrams obtained by insertion of
$\Pi_{4(2,2)}$
in $M_2$.
The number of diagrams contributing to $A_1^{(10)}$ is 3.
The contribution to $A_2^{(10)}$ comes from 21 diagrams ($3 (2^3 -1) =
21$).

\vspace{2mm}
\noindent
{\em Subset I(g)}.
Diagrams obtained by insertion of
$\Pi_{4(4)}$
in $M_2$.
The number of diagrams contributing to $A_1^{(10)}$ is  9.
The contribution to $A_2^{(10)}$ comes from 27 diagrams ($9 (2^2 -1) =
27$).

\vspace{2mm}
\noindent
{\em Subset I(h)}.
Diagrams obtained by insertion of
$\Pi_{6(2)}$
in $M_2$.
The number of diagrams contributing to $A_1^{(10)}$ is  30.
The contribution to $A_2^{(10)}$ comes from 90 diagrams ($30 (2^2 -1) =
90$).

\vspace{2mm}
\noindent
{\em Subset I(i)}.
Diagrams obtained by insertion of
$\Pi_8$ in $M_2$.
The number of diagrams contributing to $A_1^{(10)}$ is  105.
The contribution to $A_2^{(10)}$ also comes from 105 diagrams.

\vspace{2mm}
\noindent
{\em Subset I(j)}.
Diagrams obtained by insertion of eighth-order
photon propagators, which consist of two $\Lambda_4$'s
with three photons contracted, in $M_2$.
The number of diagrams contributing to $A_1^{(10)}$ is  6.
The contribution to $A_2^{(10)}$ comes from 18 diagrams ($6 (2^2 -1) =
18$).

The total number of diagrams of Set I contributing to $A_1^{(10)}$ is
208.
The number of diagrams of Set I contributing to $A_2^{(10)}$ is 498.

%%%%%%%%%%%%%%%%%%%%%%%%%%%%%%%%%%%%%%%%%%%%%%%%%%%%%%%%
\subsection{Set II}
\label{sec:set2}
%%%%%%%%%%%%%%%%%%%%%%%%%%%%%%%%%%%%%%%%%%%%%%%%%%%%%%%%%%

\begin{figure}
\includegraphics[scale=0.7]{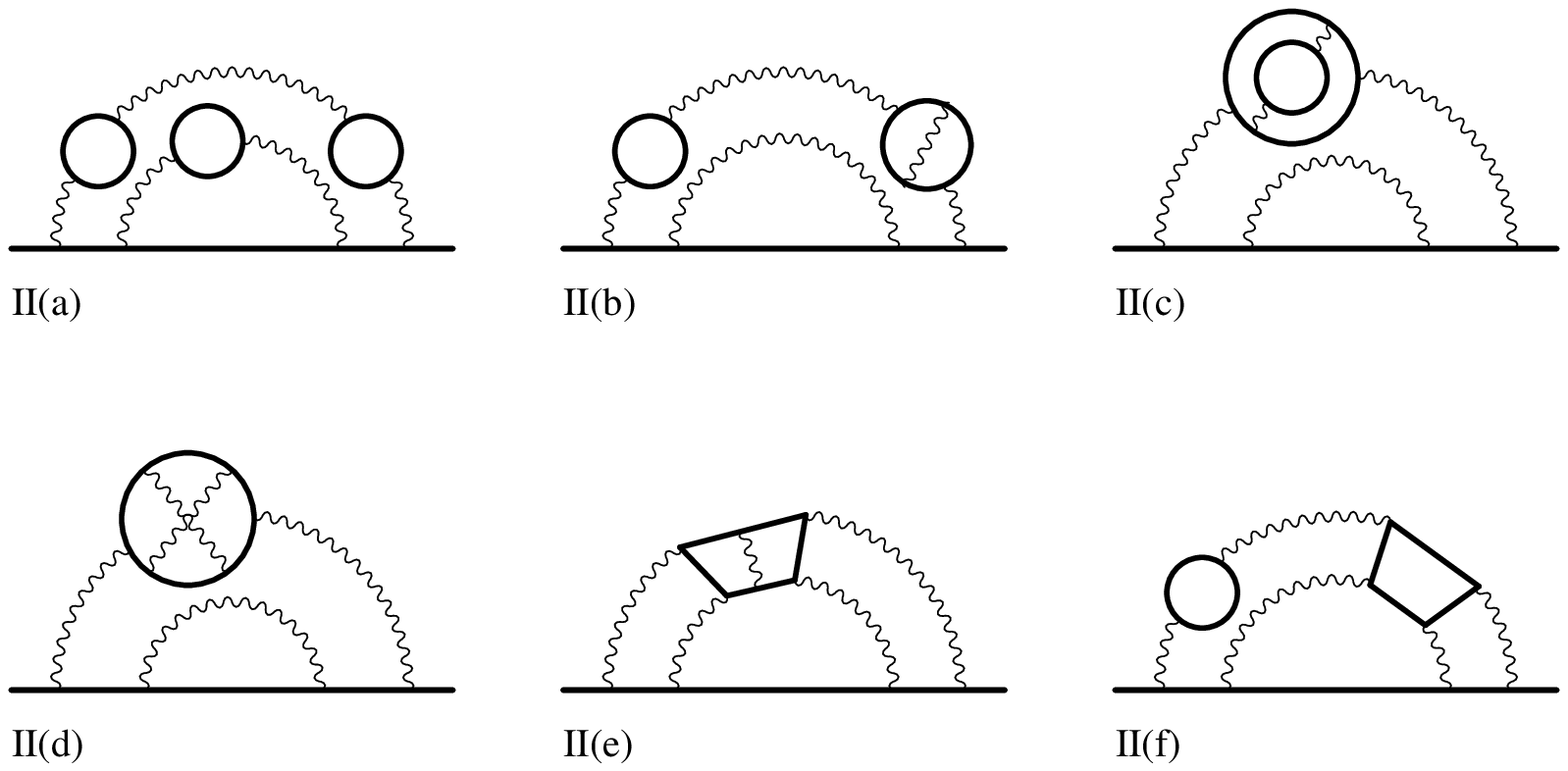}
\caption{\label{GroupII} Set II }
\end{figure}

Diagrams of Set II are built upon six proper fourth-order vertices $M_4$ 
by insertion of various closed electron loops.
It consists of 600 Feynman diagrams,
which can be classified further into six 
gauge-invariant subsets as indicated in Fig. \ref{GroupII}.

\vspace{2mm}
\noindent
{\em Subset II(a)}.
Diagrams obtained by inserting three $\Pi_2$'s
in $M_4$.
The number of diagrams contributing to $A_1^{(10)}$ is 24.
For the sake of programming convenience this set is 
subdivided into II($a_1$), in which all three $\Pi_2$'s
are inserted in the same photon line, and II($a_2$)
in which $\Pi_2$ are on both photon lines.
The total contribution to $A_2^{(10)}$ comes from 168 diagrams ($24 (2^3 -1) =
168$).

\vspace{2mm}
\noindent
{\em Subset II(b)}.
Diagrams obtained by inserting one $\Pi_2$ and one $\Pi_4$
in $M_4$.
The number of diagrams contributing to $A_1^{(10)}$ is 108.
The contribution to $A_2^{(10)}$ comes from 324 diagrams ($108 (2^2 -1)
= 324$).

\vspace{2mm}
\noindent
{\em Subset II(c)}.
Diagrams obtained by insertion of
$\Pi_{4(2)}$
in $M_4$.
The number of diagrams contributing to $A_1^{(10)}$ is 36.
The contribution to $A_2^{(10)}$ comes from 108 diagrams ($36 (2^2 -1) =
108$).

\vspace{2mm}
\noindent
{\em Subset II(d)}.
Diagrams obtained by insertion of
$\Pi_6$
in $M_4$.
The number of diagrams contributing to $A_1^{(10)}$ is 180.
The contribution to $A_2^{(10)}$ comes also from 180 diagrams.

\vspace{2mm}
\noindent
{\em Subset II(e)}.
Diagrams obtained by insertion of
internal light-by-light diagram $\Lambda_4^{(2)}$
in $M_4$.
The number of diagrams contributing to $A_1^{(10)}$ is 180.
The contribution to $A_2^{(10)}$ comes also from 180 diagrams.

\vspace{2mm}
\noindent
{\em Subset II(f)}.
Diagrams obtained by insertion of
internal light-by-light diagram $\Lambda_4$
and additional $\Pi_2$
in $M_4$.
The number of diagrams contributing to $A_1^{(10)}$ is 72.
The contribution to $A_2^{(10)}$ comes from 216 diagrams ($72 (2^2 -1) =
216$).

The total number of diagrams of Set II contributing to $A_1^{(10)}$ is
600.
The number of diagrams of Set II contributing to $A_2^{(10)}$ is 1176.

%%%%%%%%%%%%%%%%%%%%%%%%%%%%%%%%%%%%%%%%%%%%%%%%%%%%%%%%
\subsection{Set III}
\label{sec:set3}
%%%%%%%%%%%%%%%%%%%%%%%%%%%%%%%%%%%%%%%%%%%%%%%%%%%%%%%%%%

\begin{figure}
\vspace*{2cm}
\includegraphics[scale=0.7]{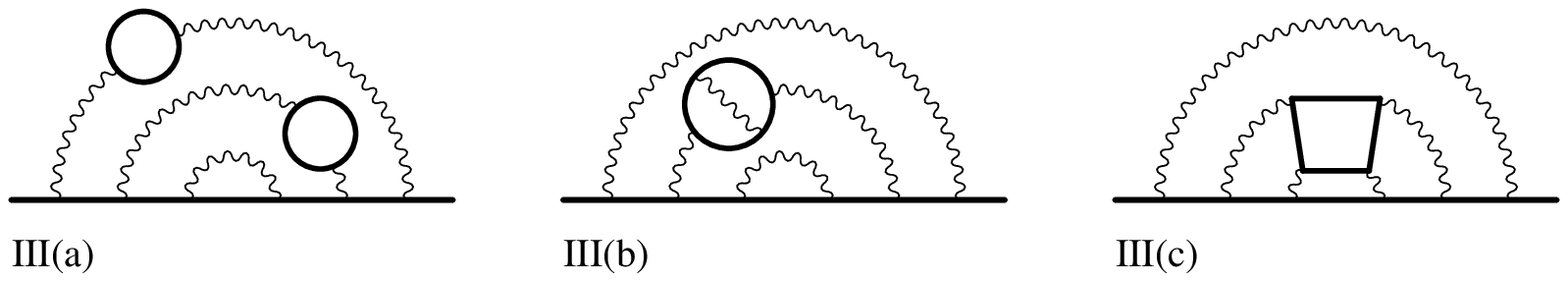}
\caption{\label{GroupIII} Set III }
\end{figure}

Diagrams of Set III are built from 50 proper sixth-order vertices $M_6$ 
by insertion of various closed electron loops.
This set consists of 1140 Feynman diagrams which
can be classified further into
three gauge-invariant subsets as indicated in Fig. \ref{GroupIII}.

\vspace{2mm}
\noindent
{\em Subset III(a)}.
Diagrams obtained by inserting two $\Pi_2$'s
in  $M_6$.
The number of diagrams contributing to $A_1^{(10)}$ is 300.
The contribution to $A_2^{(10)}$ comes from 900 diagrams ($300 (2^2 -1)
= 900$).

\vspace{2mm}
\noindent
{\em Subset III(b)}.
Diagrams obtained by inserting $\Pi_4$
in $M_6$.
The number of diagrams contributing to $A_1^{(10)}$ is 450.
The contribution to $A_2^{(10)}$ comes  also from 450 diagrams.

\vspace{2mm}
\noindent
{\em Subset III(c)}.
Diagrams obtained by insertion of
an internal light-by-light diagram $\Lambda_4$
in $M_6$.
The number of diagrams contributing to $A_1^{(10)}$ is 390.
The contribution to $A_2^{(10)}$ comes  also from 390 diagrams.

The total number of diagrams of Set III contributing to $A_1^{(10)}$ is
1140.
The number of diagrams of Set II contributing to $A_2^{(10)}$ is 1740.

%%%%%%%%%%%%%%%%%%%%%%%%%%%%%%%%%%%%%%%%%%%%%%%%%%%%%%%%
\subsection{Set IV}
\label{sec:set4}
%%%%%%%%%%%%%%%%%%%%%%%%%%%%%%%%%%%%%%%%%%%%%%%%%%%%%%%%%%

\begin{figure}
\includegraphics[scale=0.8]{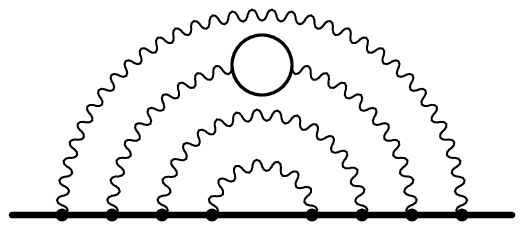}
\caption{\label{GroupIV} Set IV }
\end{figure}

\vspace{2mm}
Set IV is built from 518 proper eighth-order vertices $M_8$
by inserting a closed lepton loop of second order $\Pi_2$.
It has only one subset.
Total numbers of diagrams of Set IV contributing to $A_1^{(10)}$
and $A_2^{(10)}$ are both 2072.
A representative diagram is shown in Fig. \ref{GroupIV}.

%%%%%%%%%%%%%%%%%%%%%%%%%%%%%%%%%%%%%%%%%%%%%%%%%%%%%%%%
\subsection{Set V}
\label{sec:set5}
%%%%%%%%%%%%%%%%%%%%%%%%%%%%%%%%%%%%%%%%%%%%%%%%%%%%%%%%%%

\begin{figure}
\includegraphics[scale=0.8]{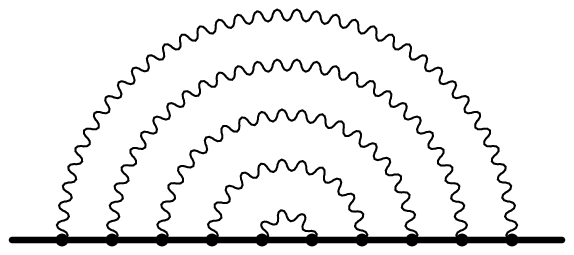}
\caption{\label{GroupV} Set V }
\end{figure}

Set V consists of proper tenth-order vertices.
It has only one subset
consisting of 6354 diagrams and contributes only to $A_1^{(10)}$.
A representative diagram is shown in Fig. \ref{GroupV}.

%%%%%%%%%%%%%%%%%%%%%%%%%%%%%%%%%%%%%%%%%%%%%%%%%%%%%%%%
\subsection{Set VI}
\label{sec:set6}
%%%%%%%%%%%%%%%%%%%%%%%%%%%%%%%%%%%%%%%%%%%%%%%%%%%%%%%%%%

\begin{figure}
\includegraphics[scale=0.7]{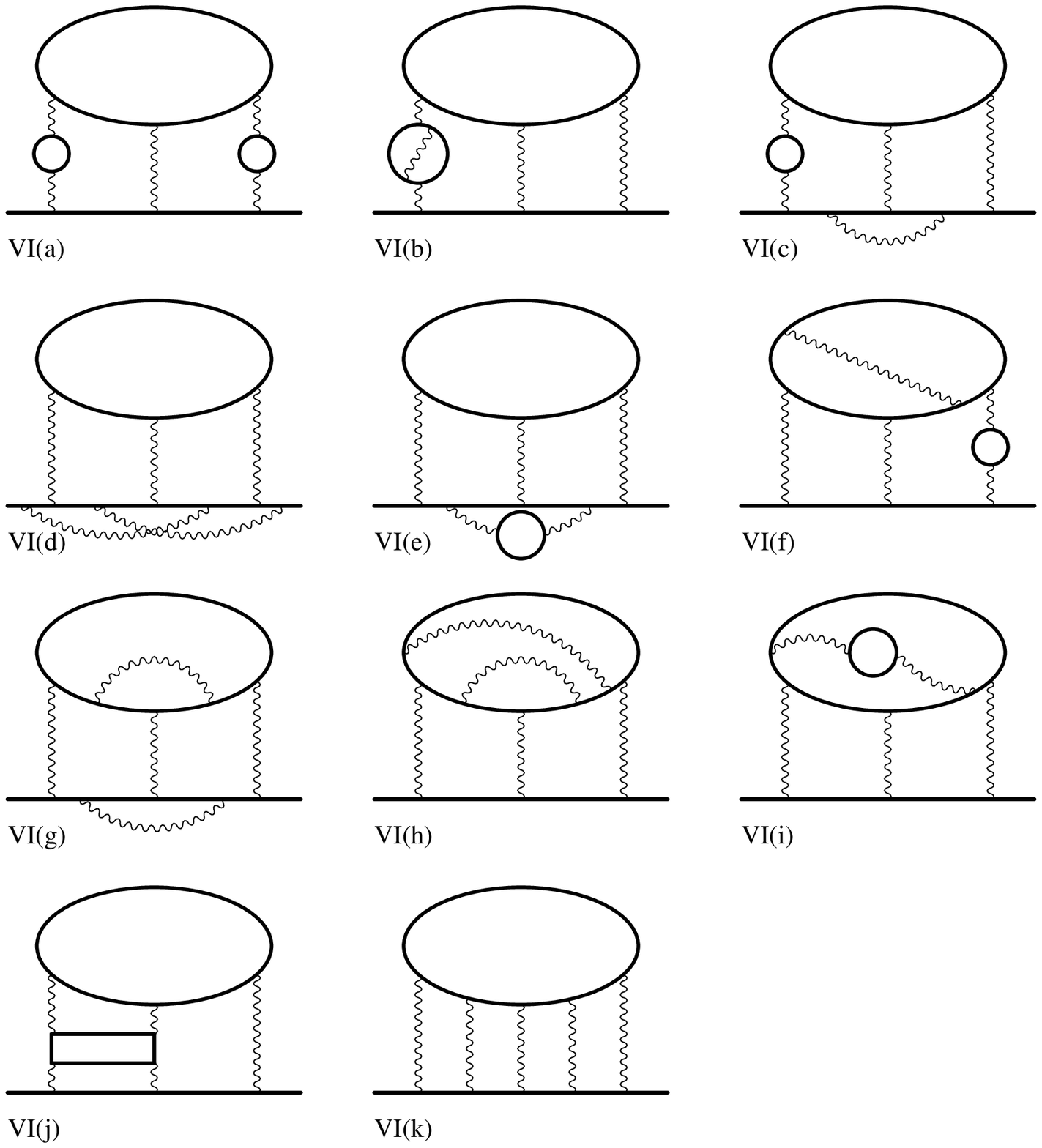}
\caption{\label{GroupVI} Set VI }
\end{figure}

This set consists of vertex diagrams of various orders
which contain at least one $l$-$l$ subdiagram.
Most diagrams of Set VI are built starting from the sixth-order diagram
$M_{6LL}$
which contains an external light-by-light-scattering subdiagram
$\Lambda_4$.
An exception is one subset that contains a subdiagram $\Lambda_6$.
Note also that diagrams 
already contained in the sets I, II, and III are excluded.
The set VI consists of 2298 Feynman diagrams which can be subdivided into
eleven gauge-invariant subsets as indicated in Fig. \ref{GroupVI}.

\vspace{2mm}
\noindent
{\em Subset VI(a)}.
Diagrams obtained by inserting two $\Pi_2$'s
in the internal photon lines of $M_{6LL}$.
The number of diagrams contributing to $A_1^{(10)}$ is 36.
The contribution to $A_2^{(10)}$ comes from 252 diagrams ($36 (2^3 -1) =
252$).

\vspace{2mm}
\noindent
{\em Subset VI(b)}.
Diagrams obtained by inserting a $\Pi_4$
in the internal photon lines of sixth-order diagram $M_{6LL}$.
The number of diagrams contributing to $A_1^{(10)}$ is 54.
The contribution to $A_2^{(10)}$ comes from 162 diagrams ($54 (2^2 -1) =
162$).

\vspace{2mm}
\noindent
{\em Subset VI(c)}.
Diagrams obtained by inserting a $\Pi_2$
in the photon lines connecting the closed lepton loop
$\Lambda_4$ and the open lepton line
in the eighth-order diagrams $M_{8LLc}$.
The number of diagrams contributing to $A_1^{(10)}$ is 144.
The contribution to $A_2^{(10)}$ comes from 432 diagrams ($144 (2^2 -1)
= 432$).

\vspace{2mm}
\noindent
{\em Subset VI(d)}.
This subset consists of diagrams
in which the open lepton line of $M_{6LL}$ is
modified by fourth-order radiative corrections.
The number of diagrams contributing to $A_1^{(10)}$ is 492.
The contribution to $A_2^{(10)}$ comes  also from 492 diagrams.

\vspace{2mm}
\noindent
{\em Subset VI(e)}.
This subset consists of diagrams
in which the open lepton line of $M_{6LL}$ is
modified by second-order radiative correction
whose photon line has $\Pi_2$ insertion.
The number of diagrams contributing to $A_1^{(10)}$ is 48.
The contribution to $A_2^{(10)}$ comes from 144 diagrams ($48 (2^2 -1) =
144$).

\vspace{2mm}
\noindent
{\em Subset VI(f)}.
This subset is derived from diagrams of $M_{8LLb}$
in which photon lines connecting $\Lambda_6$
to the open lepton line receive a $\Pi_2$ insertion.
The number of diagrams contributing to $A_1^{(10)}$ is 180.
The contribution to $A_2^{(10)}$ comes from 540 diagrams ($180 (2^2 -1)
= 540$).

\vspace{2mm}
\noindent
{\em Subset VI(g)}.
This subset consists of diagrams in which
both the closed loop $\Lambda_4$
and the open lepton line are
modified by second-order radiative corrections.
The number of diagrams contributing to $A_1^{(10)}$ is 480.
The contribution to $A_2^{(10)}$ comes also from 480 diagrams.

\vspace{2mm}
\noindent
{\em Subset VI(h)}.
This subset consists of diagrams in which
$\Lambda_4$ is modified by fourth-order radiative corrections.
The number of diagrams contributing to $A_1^{(10)}$ is 630.
The contribution to $A_2^{(10)}$ comes also from 630 diagrams.

\vspace{2mm}
\noindent
{\em Subset VI(i)}.
This subset consists of diagrams in which
$\Lambda_4$ is modified by second-order radiative correction
whose photon has a {\it v.p} insertion $\Pi_2$.
The number of diagrams contributing to $A_1^{(10)}$ is 60.
The contribution to $A_2^{(10)}$ comes from 144 diagrams ($60 (2^2 -1) =
180$).

\vspace{2mm}
\noindent
{\em Subset VI(j)}.
This subset is obtained from $M_{6LL}$
(which contains a $\Lambda_4$) by hanging
a second $\Lambda_4$ on two of three photon lines
attached to the open lepton line.
The number of diagrams contributing to $A_1^{(10)}$ is 54.
The contribution to $A_2^{(10)}$ comes from 162 diagrams ($54 (2^2 -1) =
162$).

\vspace{2mm}
\noindent
{\em Subset VI(k)}.
This subset consists of vertex diagrams containing
$\Lambda_6$ five of whose photon lines end up on the open
lepton line.
The number of diagrams contributing to $A_1^{(10)}$ is 120.
The contribution to $A_2^{(10)}$ comes  also from 120 diagrams.

The total number of diagrams of Set VI contributing to $A_1^{(10)}$ is
2298.
The number of diagrams of Set VI contributing to $A_2^{(10)}$ is 3594.

The total number of Feynman diagrams contributing to $A_1^{(10)}$
is the sum of contributions from all diagrams described above,
which is 12672.
The total number for $A_2^{(10)}$ is  9080.

The number of Feynman diagrams contributing to 
$A_3^{(10)}$ can be readily derived from the above result.

%%%%%%%%%%%%%%%%%%%%%%%%%%%%%%%%%%%%%%%%%%%%%%%%%%%%%%%%
\section{Leading diagrams contributing to $a_\mu^{(10)}$}
\label{sec:leading}
%%%%%%%%%%%%%%%%%%%%%%%%%%%%%%%%%%%%%%%%%%%%%%%%%%%%%%%%%%

Fortunately, it is not difficult to identify diagrams which
may give large contribution to $a_\mu^{(10)}$.
They are diagrams containing $\ln (m_\mu / m_e )$ terms
which tend to have large numerical values because $m_\mu$
is much larger than $m_e$.

One source of $\ln (m_\mu /m_e )$ is the vacuum-polarization
contribution to the photon propagator
which yields the logarithmic factor
as a consequence of charge renormalization.
The renormalized photon propagator has the form
\begin{equation}
D_R^{\mu ,\nu} (q) = -i \frac{g^{\mu \nu}}{q^2 } d_R (q^2 /m_e^2 , \alpha )
+ \cdots ,
\label{ren.p.p}
\end{equation}
where, to order $\alpha$,
\begin{equation}
 d_R (q^2 /m_e^2 , \alpha ) = 1 + \frac{\alpha}{\pi}
\left [ \frac{1}{3} \ln (q^2 /m_e^2 ) - \frac{5}{9} + \cdots \right ] .
\label{d_R}
\end{equation}
When $D_R$ is inserted in $g-2$ diagrams,
the momentum scale is set by the muon mass.
Thus the $\alpha/\pi$ term will give a factor of the order of
\begin{equation}
K_\eta \equiv \frac{2}{3} \ln (\eta(m_\mu /m_e) ) - \frac{5}{9}, 
\label{Kdef}
\end{equation}
where $\eta$ is an fuzzy factor of order 1.
$K_\eta \simeq 3$ for $\eta = 1$.

As a matter of fact, more important sources of $\ln (m_\mu /m_e )$
are diagrams built up from the large sixth-order
diagram $M_{6LL}$ containing an {\it external} light-by-light scattering
subdiagram of closed electron loop $\Lambda_4$.
(By {\it external} we mean that one of four photon leg is
the external magnetic field.  If all photons are virtual
photons, we call it {\it internal}.)
The extraordinary size of $M_{6LL}$ was initially discovered
by numerical integration.
The primary cause of large size is that 
it has a logarithmic mass-singularity for $m_e \rightarrow 0$.
But this is not the whole story.
It was pointed out by Yelikhoskii  \cite{yelli}
that, in the large $m_\mu /m_e$ limit,
the muon line may be regarded as a static source of
Coulomb photons as well as hyperfine spin-spin interaction.
Of three photons exchanged between the muon line and the electron loop,
one photon is responsible for the spin-spin interaction
while the other two are essentially static
Coulomb potential.
Integration over these Coulomb photon momenta gives a factor
$i\pi $ each, contributing a factor $\pi^2 (\sim 10)$
to the leading term
\begin{equation}
M_{6LL}^{(leading)} = \frac{2\pi^2}{3} \ln (m_\mu /m_e ) .
\label{leadingM6LL}
\end{equation}
For the physical value of $m_\mu /m_e$, this is about 35.
$M_{6LL}$ as a whole
is reduced by the negative mass-independent term to about 21,
which is still very large.
The value of $M_{6LL}$ containing an electron loop  $\Lambda_4$ is known exactly
and its numerical value is \cite{laporta2}
\begin{equation}
M_{6LL} = 20.947~924~34~(21),
\label{numericalM6LL}
\end{equation}
where the uncertainty is due to that of the muon mass only.

The difference between $M_{6LL}$ and $M_{6LL}^{(leading)}$
may be interpreted as an indicator of the degree to which the picture of
static Coulomb potential is valid.
One way to incorporate this is to introduce a fudge factor $\xi$
such that
\begin{equation}
M_{6LL} = \xi^2 M_{6LL}^{(leading)}.
\label{fudge}
\end{equation}
In this interpretation we have $\xi \simeq 0.77$.

These two sources of $\ln (m_\mu /m_e)$,
light-light-scattering loop and vacuum-polarization loop,
can work together and give even larger numerical factors.
For instance, the leading term of the integral $A_2 [VI(a)]$,
which has insertion of two electron loops $\Pi_2$ 
and should have been written as VI(a)[e,e,e] following the notation
of Table \ref{table2a},
will be of order \cite{oldtk1}
\begin{equation}
A_2 [VI(a)] \simeq 6 K_\eta^2 M_{6LL} ,
\label{leadingVIa}
\end{equation}
where  the factor 6 is the number of ways
two electron loops $\Pi_2$ can be inserted in $M_{6LL}$.
This leads to $A_2 [VI(a)] \simeq 1130$ for $\eta=1$ ($K_\eta \simeq 3$).
As is seen later, the actual value is about 543.
Thus $K_\eta \sim 2$.

Similarly, the subset VI(b), which has insertion of one $\Pi_4$ in $M_{6LL}$
(VI(b)[e,e] in the notation of Table \ref{table2a}),
will give contribution of the order
\begin{equation}
A_2 [VI(b)] \sim 3 \times \frac{3}{4} \times  K_\eta M_{6LL} \simeq 142,
\label{leadingVIb}
\end{equation}
for $\eta=1$ with the help of the identity \cite{karsh}
\begin{equation}
\Pi_4 (k^2)=  \frac{\alpha}{\pi} \frac{3}{4} \Pi_2 (k^2) 
+ \left(\frac{\alpha}{\pi} \right)^2 k^2 \left(\zeta (3) + \frac{5}{24}\right) .
\end{equation}
The numerical evaluation give $A_2$[VI(b)]$\simeq$ 169.
Thus $K_\eta \sim 3.5$ in this case.

If we apply blindly the argument based on (\ref{Kdef}), we would obtain,
for $K_\eta \sim 3$,
\begin{eqnarray}
A_2 [VI(c)] \simeq &3 K_\eta M_{8LLc} \simeq 27,    \nonumber   \\ 
A_2 [VI(e)] \simeq &K_\eta M_{8LLc} \simeq 9,     \nonumber   \\
A_2 [VI(f)] \simeq &3 K_\eta M_{8LLb} \simeq -4.5,     \nonumber   \\
A_2 [VI(i)] \simeq &K_\eta M_{8LLb} \simeq -1.5 .
\label{unreliable}
\end{eqnarray}
Unfortunately, these estimates are completely misleading because
$M_{8LLc}$ is the sum of large terms which tend to cancel
each other.  Similarly for $M_{8LLb}$.

Subsets VI(d), VI(g), and  VI(j) have no $\Pi_2$ insertion 
and thus will have no particular enhancement
although individual members of these subsets 
(being gauge-dependent) might have rather large values.

Among the diagrams of Sets I - V, the
term with the highest power of logarithm is 
\begin{equation}
A_2 [I(a)] \sim \frac{8}{81} \ln^4 (m_\mu /m_e ) ,
\end{equation}
although its value ($\sim 80$) is not dominating because of
its small numerical factor.
Actually terms of lower logarithmic powers reduce this 
to an even smaller value ($\sim 20$).

%Similar estimate can be made for all diagrams of Sets I - IV.
%E. g.
%%
%\begin{equation}
%A_2 [IV] \sim 4 \times K_\eta \times 1.8 \simeq 20,
%\label{leadingIV}
%\end{equation}
%%
%for $\eta=1$.
%
In the study of eighth-order terms, the value of $K_\eta$
was found  to be smaller than 3 in most cases and 
varies in the range $2 < K_\eta < 2.5$, except for VI(b).
Thus the estimates given above for $\eta=1$ 
(or $K_\eta \sim 3$) are likely to be overestimates
by $20 \sim 50 \%$.

The contribution of the subset VI(k) 
had been estimated in a different manner.
It is shown \cite{yelli} that the leading term in the large
$m_\mu/m_e$ limit is of the form
\begin{equation}
A_2 [VI(k)] = \pi^4 (0.438.. \ln (m_\mu / m_e ) + \ldots ).
\label{VIkleading}
\end{equation}
Based on this observation it was estimated that \cite{karsh}
\begin{equation}
A_2 [VI(k)] \simeq 185 \pm 85.
\label{VIkestimate}
\end{equation}
This term is large mainly because of the presence of
the $\pi4$ factor.  Its origin can be readily understood
by a generalization of the argument leading to Eq. (21)
to the case in which 2n+1 photons are exchanged between
the light-by-light type loop $\Lambda_{2n+2}$, n=1,2,...,
and the muon line [44].  In the large $m_\mu/m_e$ limit
this mechanism generates the structure 
%
%Since this is large and has a large uncertainty,
%it would be useful to see whether it can be improved in some way.
%It starts from the observation that the sixth-order contribution to
%the muon g-2 from diagrams containing a light-light scattering
%subdiagram $\Lambda_4$ may be written as that shown in (\ref{numericalM6LL}).
%
%\begin{equation}
%M_{6LL}^{leading} = \frac{2 \pi^2}{3} \ln (m_\mu/m_e) + \ldots,
%\label{l-by-l}
%\end{equation}
%
%in the large $m_\mu/m_e$ limit.
%The logarithmic term arises from the fact that the $l-l$ loop
%has a mass singularity in the small $m_e$ limit if all photon
%momenta are small compared to $m_e$.
%In this region, one of the photons exchanged between the muon
%and the electron is responsible for the hyperfine spin-spin interaction,
%but the other two act essentially like a static Coulomb potential.
%Integration over the Coulomb photon momenta gives the factor
%$i\pi$ each, contributing a factor $\pi^2 (\sim 10)$ in(\ref{l-by-l}).
%For the physical value of $m_\mu/m_e$, this is about 35.
%$M_{6LL}$ as a whole is reduced by the negative mass-independent term
%of about -15 and becomes about 20, which is still sizable.
%It is to be noted that the enhancement factor $\pi^2$ may not be
%present for the mass-independent term.
%
%%The argument leading to (\ref{numericalM6LL}) can be readily
%generalized to the case in which
%$2n+1$ photons are exchanged between the light-by-light type loop
%$\Lambda_{2n+2}$, $n = 1, 2, ...$, and the muon line \cite{yelli}).
%%%In the large $m_\mu/m_e$ limit one has the structure
%
\begin{equation}
A_2^{(2n+1)}  \simeq c_n \pi^{2n} \ln (m_\mu/m_e) +  \cdots .
\end{equation}
Numerical values of some $c_n$ are known \cite{milstein}
\begin{eqnarray}
c_1 &=& \frac{2}{3},   \nonumber   \\
c_2 &=& 0.438 \ldots .
\end{eqnarray}
%

%%%%%%%%%%%%%%%%%%%%%%%%%%%%%%%%%%%%%%%%%%%%%%%%%%%%%%%%
\section{Analytically known contribution to $a_\mu^{(10)}$}
\label{sec:analytic}
%%%%%%%%%%%%%%%%%%%%%%%%%%%%%%%%%%%%%%%%%%%%%%%%%%%%%%%%%%

At present only a small number of integrals in the
subsets I(a), I(b), I(c), II(a), and II(b)
are known analytically.  Their expansion in the ratio $m_e/m_\mu$
are given in \cite{laportaX}.
From Table 2 of \cite{laportaX} we obtain
\begin{eqnarray}
a_\mu [I(a)] &=& ~22.566~973~(3),  \nonumber  \\
a_\mu [I(b)] &=& ~30.667~091~(3),  \nonumber  \\
a_\mu [I(c)] &=& ~~5.141~395~(1),  \nonumber  \\
a_\mu [II(a)] &=&-36.174~859~(2),  \nonumber  \\
a_\mu [II(b)] &=&-23.462~173~(1),
\label{anal_result}
\end{eqnarray}
where the uncertainties come from the measurement uncertainty of
$m_e/m_\mu$ only.
These results show strong cancellation among diagrams
of Sets I and II, which is analogous with the cancellation
among Group I and Group II diagrams contributing to
the eighth-order term of muon g-2.

%%%%%%%%%%%%%%%%%%%%%%%%%%%%%%%%%%%%%%%%%%%%%%%%%%%%%%%%
\section{Numerical evaluation of diagrams contributing to $a_\mu^{(10)}$}
\label{sec:numerical}
%%%%%%%%%%%%%%%%%%%%%%%%%%%%%%%%%%%%%%%%%%%%%%%%%%%%%%%%%%

As was discussed in Section \ref{sec:classification}
diagrams containing light-by-light-scattering
subdiagrams, namely the Set VI, is the source of dominant contribution to $a_\mu^{(10)}$.
However, let us describe the results of numerical evaluation
starting from the Set I.

%%%%%%%%%%%%%%%%%%%%%%%%%%%%%%%%%%%%%%%%%%%%%%%%%%%%%%%%
\subsection{Set I}
\label{sec:setone}
%%%%%%%%%%%%%%%%%%%%%%%%%%%%%%%%%%%%%%%%%%%%%%%%%%%%%%%%%%

Integrals for the diagrams of Set I, except for subsets I(i)
and I(j), can be readily obtained from eighth-order diagrams
belonging to Group I by insertion of {\it v-p} loops.
Diagrams numerically evaluated thus far are 
shown in Tables \ref{table1}, \ref{table1d} and \ref{table1f}.
Summing up the first four lines of Table \ref{table1}
one obtains
\begin{eqnarray}
A_2 [I(a)] &=& 
M_{2,p2:4}^{(e,e,e,e)}
+M_{2,p2:4}^{(e,e,e,m)}
+M_{2,p2:4}^{(e,e,m,m)}
+M_{2,p2:4}^{(e,m,m,m)}   \nonumber  \\
&=& 22.567~05~(25).
\label{A2Ia}
\end{eqnarray}
Similarly, we find from the rest of Table \ref{table1}
\begin{eqnarray}
A_2 [I(b)] &=& 
M_{2,p4,p2:2}^{(e,e,e)}
+M_{2,p4,p2:2}^{(e,m,e)}
+M_{2,p4,p2:2}^{(e,m,m)}
+M_{2,p4,p2:2}^{(m,e,e)}
+M_{2,p4,p2:2}^{(m,m,e)}
\nonumber  \\
&=& 30.667~54~(33),
\nonumber \\
A_2 [I(c)] &=& 
M_{2,p4:2}^{(e,e)}
+M_{2,p4:2}^{(e,m)}
\nonumber  \\
&=& 5.141~38~(15),
\label{A2Ib-c}
\end{eqnarray}

The values of 
$A_2 [I(a)]$, $A_2 [I(b)]$, and $A_2 [I(c)]$ are in good
agreement with semi-analytic values  \cite{laportaX}
quoted in (\ref{anal_result}).

$A_2 [I(d)]$ is evaluated from the entries in Table \ref{table1d}
together with Table III and Table VI of \cite{kn4},
which list residual renormalization terms. 
We obtain
\begin{eqnarray}
M_{2,p4(p2)p2}^{(e(e),e)}
&=&2 M_{2,p4A(p2)p2}^{(e(e),e)}
+  4 M_{2,p4B(p2)p2}^{(e(e),e)}
-  4 \Delta B_{2,p2}^{(e,e)} M_{2,p2:2}^{(m,e,e)}   \nonumber   \\
&=& 7.45173~(101),
\nonumber   \\
M_{2,p4(p2)p2}^{(m(e),e)}
&=&2 M_{2,p4A(p2)p2}^{(m(e),e)}
+  4 M_{2,p4B(p2)p2}^{(m(e),e)}
-  4 \Delta B_{2,p2}^{(m,e)} M_{2,p2:2}^{(m,m,e)}   \nonumber   \\
&=& 1.02576~(12),
\nonumber   \\
M_{2,p4(p2)p2}^{(e(m),e)}
&=&2 M_{2,p4A(p2)p2}^{(e(m),e)}
+  4 M_{2,p4B(p2)p2}^{(e(m),e)}
-  4 \Delta B_{2,p2}^{(e,m)} M_{2,p2:2}^{(m,e,e)}   \nonumber   \\
&=& 0.13084~(2),
\nonumber   \\
M_{2,p4(p2)p2}^{(m(m),e)}
&=&2 M_{2,p4A(p2)p2}^{(m(m),e)}
+  4 M_{2,p4B(p2)p2}^{(m(m),e)}
-  4 \Delta B_{2,p2}^{(m,m)} M_{2,p2:2}^{(m,m,e)}   \nonumber   \\
&=& 7.17353~(81) \times 10^{-2},
\label{A2Id_part1}
\end{eqnarray}
\begin{eqnarray}
M_{2,p4(p2)p2}^{(e(e),m)}
&=&2 M_{2,p4A(p2)p2}^{(e(e),m)}
+  4 M_{2,p4B(p2)p2}^{(e(e),m)}
-  4 \Delta B_{2,p2}^{(e,e)} M_{2,p2:2}^{(m,m,e)}   \nonumber   \\
&=& 0.15845~(3),
\nonumber   \\
M_{2,p4(p2)p2}^{(m(e),m)}
&=&2 M_{2,p4A(p2)p2}^{(m(e),m)}
+  4 M_{2,p4B(p2)p2}^{(m(e),m)}
-  4 \Delta B_{2,p2}^{(m,e)} M_{2,p2:2}^{(m,m,m)}   \nonumber   \\
&=& 4.82662~(1) \times 10^{-2},
\nonumber   \\
M_{2,p4(p2)p2}^{(e(m),m)}
&=&2 M_{2,p4A(p2)p2}^{(e(m),m)}
+  4 M_{2,p4B(p2)p2}^{(e(m),m)}
-  4 \Delta B_{2,p2}^{(e,m)} M_{2,p2:2}^{(m,e,m)}   \nonumber   \\
&=& 5.29020~(63)\times 10^{-3},
\label{A2Id_part2}
\end{eqnarray}
From these contributions we obtain
$A_2 [I(d)]$:
\begin{eqnarray}
A_2 [I(d)] &=& 
M_{2,p4(p2)p2}^{(e(e),e)}
+M_{2,p4(p2)p2}^{(m(e),e)}
+M_{2,p4(p2)p2}^{(e(m),e)}
+M_{2,p4(p2)p2}^{(m(m),e)}  \nonumber  \\
&+&M_{2,p4(p2)p2}^{(e(e),m)}
+M_{2,p4(p2)p2}^{(m(e),m)}
+M_{2,p4(p2)p2}^{(e(m),m)}
\nonumber  \\
&=& 8.89207~(102).
\label{A2Id}
\end{eqnarray}
Several terms of $A_2 [I(d)]$ was also evaluated by an alternative method using 
an exact spectral function of $\Pi_4 (\Pi_2)$ in which two loop masses are
equal.  They are listed at the bottom of Table \ref{table1d}.
They are in good agreement with the values given in 
(\ref{A2Id_part1}) and (\ref{A2Id_part2}).

$A_2 [I(e)]$ is evaluated using the Pad\'{e} approximant for $\Pi_6$,
which is known to be a very good approximation
from an earlier work. Our result is

\begin{eqnarray}
A_2 [I(e)] &=& 
M_{2,p6p2}^{(e,e)}
+M_{2,p6p2}^{(e,m)}
+M_{2,p6p2}^{(m,e)}
\nonumber  \\
&=& -1.219~20~(71).
\label{A2Ie}
\end{eqnarray}

$A_2 [I(f)]$ is evaluated using the entries of Table \ref{table1f} together
with Table \ref{table3}.  We obtain
\begin{eqnarray}
a_{2,p4(p2:2)}^{(m,e(e,e))}
&=&\Delta M_{2,p4A(p2:2)}^{(m,e(e,e))}
+\Delta M_{2,p4B(p2:2)}^{(m,e(e,e))}
-  2 \Delta B_{2,p2:2}^{(e(e,e))} M_{2,p2}^{(m,e)}   \nonumber   \\
&=& 2.88598~(9),
\nonumber   \\
a_{2,p4(p2:2)}^{(m,e(e,m))}
&=&\Delta M_{2,p4A(p2:2)}^{(m,e(e,m))}
+\Delta M_{2,p4B(p2:2)}^{(m,e(e,m))}
-  2 \Delta B_{2,p2:2}^{(e(e,m))} M_{2,p2}^{(m,e)}   \nonumber   \\
&=& 0.16111~(3),
\nonumber   \\
a_{2,p4(p2:2)}^{(m,e(m,m))}
&=&\Delta M_{2,p4A(p2:2)}^{(m,e(m,m))}
+\Delta M_{2,p4B(p2:2)}^{(m,e(m,m))}
-  2 \Delta B_{2,p2:2}^{(e(m,m))} M_{2,p2}^{(m,e)}   \nonumber   \\
&=& 0.01063~(1),
\nonumber   \\
a_{2,p4(p2:2)}^{(m,m(e,e))}
&=&\Delta M_{2,p4A(p2:2)}^{(m,m(e,e))}
+\Delta M_{2,p4B(p2:2)}^{(m,m(e,e))}
-  2 \Delta B_{2,p2:2}^{(m(e,e))} M_{2,p2}^{(m,m)}   \nonumber   \\
&=& 0.53660~(9),
\nonumber   \\
a_{2,p4(p2:2)}^{(m,m(e,m))}
&=&\Delta M_{2,p4A(p2:2)}^{(m,m(e,m))}
+\Delta M_{2,p4B(p2:2)}^{(m,m(e,m))}
-  2 \Delta B_{2,p2:2}^{(m(e,m))} M_{2,p2}^{(m,m)}   \nonumber   \\
&=& 0.09079~(2),
\label{A2If_parts}
\end{eqnarray}
From these contributions we obtain
$A_2 [I(f)]$:
\begin{eqnarray}
A_2 [I(f)] &=& 
a_{2,p4(p2:2)}^{(m,e(e,e))}
+ a_{2,p4(p2:2)}^{(m,e(e,m))}
+ a_{2,p4(p2:2)}^{(m,e(m,m))}
+ a_{2,p4(p2:2)}^{(m,m(e,e))}
+ a_{2,p4(p2:2)}^{(m,m(e,m))}
\nonumber  \\
&=& 3.68510~(13).
\label{A2If}
\end{eqnarray}
%

%%%%%%%%%%%%%%%%%%%%%%%%%%%%%%%%%%%%%%%%%%%%%%%%%%%%%%%%
\subsection{Set II}
\label{sec:settwo}
%%%%%%%%%%%%%%%%%%%%%%%%%%%%%%%%%%%%%%%%%%%%%%%%%%%%%%%%%%

Integrals of Set II, except for subset II(e), 
can be readily obtained from eighth-order diagrams
belonging to Group II by insertion of {\it v-p} loops.
We have evaluated only subsets II(a),  II(b), and II(f) thus far.
From Table \ref{table2a} and Table \ref{table3}
of this paper and Tables III and VI of \cite{kn4},
we obtain
\begin{eqnarray}
M_{4,p2:3}^{(e,e,e)} &=&
2 M_{4a,p2:3}^{(e,e,e)} 
+ M_{4b,p2:3}^{(e,e,e)\alpha} 
+ M_{4b,p2:3}^{(e,e,e)\beta} 
-  \Delta B_2 M_{2,p2:3}^{(e,e,e)}
-  \Delta B_{2,p2:3}^{(e,e,e)} M_2    \nonumber   \\
&=& -28.43132~(344),
\nonumber   \\
M_{4,p2:2,p2}^{(e,e)(e)} &=&
2 M_{4a,p2:2,p2}^{(e,e)(e)} 
+ M_{4b,p2:2,p2}^{(e,e)(e)} 
+ M_{4b,p2,p2:2}^{(e)(e,e)} 
-  \Delta B_{2,p2}^{(e)} M_{2,p2:2}^{(e,e)}
-  \Delta B_{2,p2:2}^{(e,e)} M_{2,p2}^{(e)}    \nonumber   \\
&=& -27.42432~(127),
\nonumber   \\
M_{4,p2:3}^{(e,e,m)} &=&
2 M_{4a,p2:3}^{(e,e,m)()} 
+ M_{4b,p2:3}^{(e,e,m)()} 
+ M_{4b,p2:3}^{()(e,e,m)} 
-  \Delta B_2 M_{2,p2:3}^{(e,e,m)}
-  \Delta B_{2,p2:3}^{(e,e,m)} M_2   \nonumber   \\
&=& -6.79245~(56),
\nonumber   \\
M_{4,p2:2,p2}^{(e,e)(m)} &=&
2 M_{4a,p2:2,p2}^{(e,e)(m)} 
+ M_{4b,p2:2,p2}^{(e,e)(m)} 
+ M_{4b,p2,p2:2}^{(m)(e,e)} 
-  \Delta B_{2,p2}^{(m)} M_{2,p2:2}^{(e,e)}
-  \Delta B_{2,p2:2}^{(e,e)} M_{2,p2}^{(m)}    \nonumber   \\
&=& -1.95703~(37),
\label{A2IIa_part1}
\end{eqnarray}
\begin{eqnarray}
M_{4,p2:2,p2}^{(e,m)(e)} &=&
4 M_{4a,p2:2,p2}^{(e,m)(e)} 
+2 M_{4b,p2:2,p2}^{(e,m)(e)} 
+2 M_{4b,p2,p2:2}^{(e)(e,m)} 
-2 \Delta B_{2,p2}^{(e)} M_{2,p2:2}^{(e,m)}
-2 \Delta B_{2,p2:2}^{(e,m)} M_{2,p2}^{(e)}    \nonumber   \\
&=& -4.14893 (48),
\nonumber   \\
M_{4,p2:3}^{(e,m,m)} &=&
2 M_{4a,p2:3}^{(e,m,m)} 
+ M_{4b,p2:3}^{(e,m,m)()} 
+ M_{4b,p2:3}^{()(e,m,m)} 
-  \Delta B_2 M_{2,p2:3}^{(e,m,m)}
-  \Delta B_{2,p2:3}^{(e,m,m)} M_2   \nonumber   \\
&=& -0.95284~(14),
\nonumber   \\
M_{4,p2:2,p2}^{(m,m)(e)} &=&
2 M_{4a,p2:2,p2}^{(m,m)(e)} 
+ M_{4b,p2:2,p2}^{(m,m)(e)} 
+ M_{4b,p2,p2:2}^{(e)(m,m)} 
-  \Delta B_{2,p2}^{(e)} M_{2,p2:2}^{(m,m)}
-  \Delta B_{2,p2:2}^{(m,m)} M_{2,p2}^{(e)}    \nonumber   \\
&=& -0.28902~(29),
\nonumber   \\
M_{4,p2:2,p2}^{(e,m)(m)} &=&
4 M_{4a,p2:2,p2}^{(e,m)(m)} 
+2 M_{4b,p2:2,p2}^{(e,m)(m)} 
+2 M_{4b,p2,p2:2}^{(m)(e,m)} 
-2 \Delta B_{2,p2}^{(m)} M_{2,p2:2}^{(e,m)}
-2 \Delta B_{2,p2:2}^{(e,m)} M_{2,p2}^{(m)}    \nonumber   \\
&=& -0.47576 (23).
\label{A2IIa_part2}
\end{eqnarray}
Adding up these contributions one obtains
\begin{equation}
A_2 [II(a)] = -70.4717~(38).
\label{A2IIa}
\end{equation}

From Table  \ref{table2b} and Table \ref{table3}
of this paper and Tables III and VI of \cite{kn4} we obtain
\begin{eqnarray}
M_{4,p4p2}^{(e,e)} &=&
2\Delta_1 M_{4a,p4p2}^{(e,e)()} 
+\Delta_1 M_{4b,p4p2}^{(e,e)()} 
+\Delta_1 M_{4b,p4p2}^{()(e,e)}   
+(p4 \leftrightarrow p2)  \nonumber   \\
&-&  \Delta B_2 M_{2,p4p2}^{(e,e)}
-  \Delta B_{2,p4p2}^{(e,e)} M_2  
\nonumber   \\
&=& -19.04191~(203),
\nonumber   \\
M_{4,p4p2}^{(e,m)} &=&
2\Delta_1 M_{4a,p4p2}^{(e,m)()} 
+\Delta_1 M_{4b,p4p2}^{(e,m)()} 
+\Delta_1 M_{4b,p4p2}^{()(e,m)}   
+(p4 \leftrightarrow p2)  \nonumber   \\
&-&  \Delta B_2 M_{2,p4p2}^{(e,m)}
-  \Delta B_{2,p4p2}^{(e,m)} M_2  
\nonumber   \\
&=& -1.28845~(25),
\nonumber   \\
M_{4,p4p2}^{(m,e)} &=&
2\Delta_1 M_{4a,p4p2}^{(m,e)()} 
+\Delta_1 M_{4b,p4p2}^{(m,e)()} 
+\Delta_1 M_{4b,p4p2}^{()(m,e)}    
+(p4 \leftrightarrow p2)  \nonumber   \\
&-&  \Delta B_2 M_{2,p4p2}^{(m,e)}
-  \Delta B_{2,p4p2}^{(m,e)} M_2  
\nonumber   \\
&=& -3.13132~(25),
\label{A2IIb_part1}
\end{eqnarray}
\begin{eqnarray}
M_{4,p4p2}^{(e)(e)} &=&
2\Delta_2 M_{4a,p4p2}^{(e)(e)} 
+\Delta_2 M_{4b,p4p2}^{(e)(e)} 
  +(p4 \leftrightarrow p2)  \nonumber   \\
&-&  \Delta B_{2,p2}^{(e)} M_{2,p4}^{(e)}
-  \Delta B_{2,p4}^{(e)} M_{2,p2}^{(e)}
\nonumber   \\
&=& -9.42667~(145),
\nonumber   \\
M_{4,p4p2}^{(e)(m)} &=&
2\Delta_2 M_{4a,p4p2}^{(e)(m)} 
+\Delta_2 M_{4b,p4p2}^{(e)(m)} 
  +(p4 \leftrightarrow p2)  \nonumber   \\
&-&  \Delta B_{2,p2}^{(m)} M_{2,p4}^{(e)}
-  \Delta B_{2,p4}^{(e)} M_{2,p2}^{(m)}
\nonumber   \\
&=& -0.60877~(18),
\nonumber   \\
M_{4,p4p2}^{(m)(e)} &=&
2\Delta_2 M_{4a,p4p2}^{(m)(e)} 
+\Delta_2 M_{4b,p4p2}^{(m)(e)} 
  +(p4 \leftrightarrow p2)  \nonumber   \\
&-&  \Delta B_{2,p2}^{(e)} M_{2,p4}^{(m)}
-  \Delta B_{2,p4}^{(m)} M_{2,p2}^{(e)}
\nonumber   \\
&=& -1.27436~(27).
\label{A2IIb_parts}
\end{eqnarray}
Adding up these values we obtain
\begin{equation}
A_2 [II(b)] = -34.7715~(26).
\label{A2IIb}
\end{equation}
%
%These results lead to -36.17661(347) and -23.46168(206) for the values of
%$A_2$[II(a)] and $A_2$[II(b)] which are in good agreements 
%with the corresponding analytic values in (\ref{anal_result}).
%[More diagrams of Set II, III, and IV will be evaluated numerically.]
 
Table \ref{table2f} lists numerical results of set II(f)
obtained in Version A.  The results obtained 
by an alternate formulation (Version B)
are listed in Table \ref{table2fB}. 
Combining the values in these tables statistically one obtains
\begin{eqnarray}
A_2 [II(f)]^{(e,e)} &=& -57.0633 (109),    \nonumber    \\
A_2 [II(f)]^{(e,m)} &=&~ -4.7157 (~31),    \nonumber    \\
A_2 [II(f)]^{(m,e)} &=& -15.6857 (~37).
\label{A2IIfall}
\end{eqnarray}
From (\ref{A2IIfall}) one obtains
\begin{equation}
A_2 [II(f)] = -77.4648 (120).
\label{A2IIf}
\end{equation}
%
%

%%%%%%%%%%%%%%%%%%%%%%%%%%%%%%%%%%%%%%%%%%%%%%%%%%%%%%%%
\subsection{Set VI}
\label{sec:setsix}
%%%%%%%%%%%%%%%%%%%%%%%%%%%%%%%%%%%%%%%%%%%%%%%%%%%%%%%%%%

By far the largest contribution to $A_2^{(10)}$ comes from the subset VI(a),
followed by the subset VI(b).
Their integrals can be readily obtained by insertion of {v-p}
functions $\Pi_2$ and $\Pi_4$ into the sixth-order diagram
$M_{6LL}$. 
We have evaluated them precisely by VEGAS.
The results are listed in Table \ref{table6}. Summing them up, we obtain
\begin{equation}
A_2 [VI(a)] = 629.1407~(118),
\label{A2VIa}
\end{equation}
and
\begin{equation}
A_2 [VI(b)] = 181.1285~(51) .  
\label{A2VIb}
\end{equation}
The difference between VI(a)[e,e,e] of Table \ref{table6} and old one
in Eq.(2.49) of  \cite{oldtk1} is due to a program error in the latter.
The results in Table \ref{table6} 
show that the leading term estimate (\ref{leadingVIa})
is an overestimate by a factor two, while 
the estimate (\ref{leadingVIb}) is not too far off.

Contributions of subsets VI(c), VI(e), VI(f) and VI(i) are 
also not difficult to evaluate, since they can be readily obtained
by insertion of $\Pi_2$ into some eighth-order integrals.
Their numerical values are
listed in Tables \ref{table6c}, \ref{table6e}, \ref{table6f},
and \ref{table6i}.
From this we obtain
\begin{eqnarray}
A_2 [VI(c)]^{(e,e)} &=&   
\Delta M_{8LLEp}^{(e,e)} +
\Delta M_{8LLFp}^{(e,e)} +
\Delta M_{8LLGp}^{(e,e)} +
\Delta M_{8LLHp}^{(e,e)} + 
\Delta M_{8LLIp}^{(e,e)} 
-2\Delta B_2 M_{6LLp}^{(e,e)}
\nonumber  \\
&=& -17.0505 ~(1122),
\nonumber  \\
A_2 [VI(c)]^{(m,e)} &=&   
\Delta M_{8LLEp}^{(m,e)} +
\Delta M_{8LLFp}^{(m,e)} +
\Delta M_{8LLGp}^{(m,e)} +
\Delta M_{8LLHp}^{(m,e)} + 
\Delta M_{8LLIp}^{(m,e)} 
-2\Delta B_2 M_{6LLp}^{(m,e)}
\nonumber  \\
&=& -14.2744 ~(105),
\nonumber  \\
A_2 [VI(c)]^{(e,m)} &=&   
\Delta M_{8LLEp}^{(e,m)} +
\Delta M_{8LLFp}^{(e,m)} +
\Delta M_{8LLGp}^{(e,m)} +
\Delta M_{8LLHp}^{(e,m)} + 
\Delta M_{8LLIp}^{(e,m)} 
-2\Delta B_2 M_{6LLp}^{(e,m)}
\nonumber  \\
&=& -5.2514 ~(129),
\nonumber  \\
A_2 [VI(c)] &=&   
A_2 [VI(c)]^{(e,e)} +   
A_2 [VI(c)]^{(m,e)} +   
A_2 [VI(c)]^{(e,m)}    
\nonumber  \\
&=& -36.5763 ~(1141),
\label{6c}
\end{eqnarray}
\begin{eqnarray}
A_2 [VI(e)]^{(e,e)}&=&   
\Delta M_{8LLEq}^{(e,e)} +
\Delta M_{8LLFq}^{(e,e)} +
\Delta M_{8LLGq}^{(e,e)} +
\Delta M_{8LLHq}^{(e,e)} +
\Delta M_{8LLIq}^{(e,e)} 
\nonumber  \\
&-&2\Delta B_{2,p2}^{(m,e)} M_{6LL}
\nonumber  \\
&=& 0.7524 ~(1338),
\nonumber  \\
A_2 [VI(e)]^{(m,e)}&=&   
\Delta M_{8LLEq}^{(m,e)} +
\Delta M_{8LLFq}^{(m,e)} +
\Delta M_{8LLGq}^{(m,e)} +
\Delta M_{8LLHq}^{(m,e)} +
\Delta M_{8LLIq}^{(m,e)} 
\nonumber  \\
&-&2\Delta B_{2,p2}^{(m,e)} M_{6LL}^{(m,m)}
\nonumber  \\
&=& -3.9789~(78),
\nonumber  \\
A_2 [VI(e)]^{(e,m)}&=&   
\Delta M_{8LLEq}^{(e,m)} +
\Delta M_{8LLFq}^{(e,m)} +
\Delta M_{8LLGq}^{(e,m)} +
\Delta M_{8LLHq}^{(e,m)} +
\Delta M_{8LLIq}^{(e,m)} 
\nonumber  \\
&-&2\Delta B_{2,p2}^{(m,m)} M_{6LL}
\nonumber  \\
&=& -1.0950 ~(35),
\nonumber  \\
A_2 [VI(e)]&=&   
A_2 [VI(e)]^{(e,e)} +   
A_2 [VI(e)]^{(m,e)} +   
A_2 [VI(e)]^{(e,m)}   
\nonumber  \\
&=& -4.3215 ~(1341),
\label{6e}
\end{eqnarray}
\begin{eqnarray}
A_2 [VI(f)]^{(e,e)} &=&   
\Delta M_{8LLAp}^{(e,e)} +
\Delta M_{8LLBp}^{(e,e)} +
\Delta M_{8LLCp}^{(e,e)} +
\Delta M_{8LLDp}^{(e,e)} -3\Delta B_2 M_{6LLp}^{(e,e)}
\nonumber  \\
&=& -45.0425 ~(1463),
\nonumber  \\
A_2 [VI(f)]^{(m,e)} &=&   
\Delta M_{8LLAp}^{(m,e)} +
\Delta M_{8LLBp}^{(m,e)} +
\Delta M_{8LLCp}^{(m,e)} +
\Delta M_{8LLDp}^{(m,e)} -3\Delta B_2 M_{6LLp}^{(m,e)}
\nonumber  \\
&=& 8.7673~(228),
\nonumber  \\
A_2 [VI(f)]^{(e,m)} &=&   
\Delta M_{8LLAp}^{(e,m)} +
\Delta M_{8LLBp}^{(e,m)} +
\Delta M_{8LLCp}^{(e,m)} +
\Delta M_{8LLDp}^{(e,m)} -3\Delta B_2 M_{6LLp}^{(e,m)}
\nonumber  \\
&=& -1.8847 ~(155),
\nonumber  \\
A_2 [VI(f)] &=&   
A_2 [VI(f)]^{(e,e)} +   
A_2 [VI(f)]^{(m,e)} +   
A_2 [VI(f)]^{(e,m)}    
\nonumber  \\
&=& -38.1598 ~(1488),
\label{6f}
\end{eqnarray}
\begin{eqnarray}
A_2 [VI(i)]^{(e,e)} &=&   
\Delta M_{8LLAq}^{(e,e)} +
\Delta M_{8LLBq}^{(e,e)} +
\Delta M_{8LLCq}^{(e,e)} +
\Delta M_{8LLDq}^{(e,e)} -3\Delta B_{2,p2}^{(e,e)}  M_{6LL}
\nonumber  \\
&=& -28.3367 ~(1142),
\nonumber  \\
A_2 [VI(i)]^{(m,e)} &=&   
\Delta M_{8LLAq}^{(m,e)} +
\Delta M_{8LLBq}^{(m,e)} +
\Delta M_{8LLCq}^{(m,e)} +
\Delta M_{8LLDq}^{(m,e)} -3\Delta B_{2,p2}^{(m,e)}  M_{6LL}^{(m,m)}
\nonumber  \\
&=& 2.0977~(105),,
\nonumber  \\
A_2 [VI(i)]^{(e,m)} &=&   
\Delta M_{8LLAq}^{(e,m)} +
\Delta M_{8LLBq}^{(e,m)} +
\Delta M_{8LLCq}^{(e,m)} +
\Delta M_{8LLDq}^{(e,m)} -3\Delta B_{2,p2}^{(e,m)}  M_{6LL}
\nonumber  \\
&=& -1.0983 ~(31),
\nonumber  \\
A_2 [VI(i)] &=&   
A_2 [VI(i)]^{(e,e)} +   
A_2 [VI(i)]^{(m,e)} +   
A_2 [VI(i)]^{(e,m)}   
\nonumber  \\
&=& -27.3373 ~(1147).
\label{6i}
\end{eqnarray}

The last columns of Table \ref{table6c} and Table \ref{table6f}
show that the factor $r$ is roughly equal to $3K_\eta$ 
with $K_\eta \simeq 2$ for sets VI(c) and VI(f).
These factors vary from diagram to diagram within
each set, making naive estimates for the sums of diagrams given in
(\ref{unreliable}) entirely different from the corresponding terms
of (\ref{6c}) - (\ref{6i}) obtained by explicit numerical integration. 
The "enhancement" factor is $K_\eta \sim 1 $ for sets VI(e) and VI(i).
Table \ref{table6e} and Table \ref{table6i} show that $K_\eta \sim 1$
in these cases,
indicating that there is no significant enhancement.
This is probably
because $\Pi_2$ is buried deep in other subdiagrams in these cases.

Among the remaining subsets the subset VI(k) is 
likely to be the largest and most
challenging because it has a subdiagram $\Lambda_6$
which consists of 120 proper lepton loops to which
six photon lines are attached.
See Fig. 8.
This is a diagram which appears for the first
time in the tenth-order so that it cannot be derived
from or related to lower-order diagrams.

\begin{figure}
\includegraphics[scale=0.6]{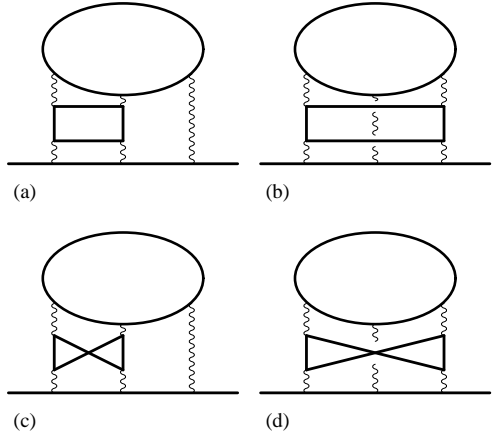}
\caption{\label{X6jall} Subset VI[j] }
\end{figure}

\begin{figure}
\includegraphics[scale=0.5]{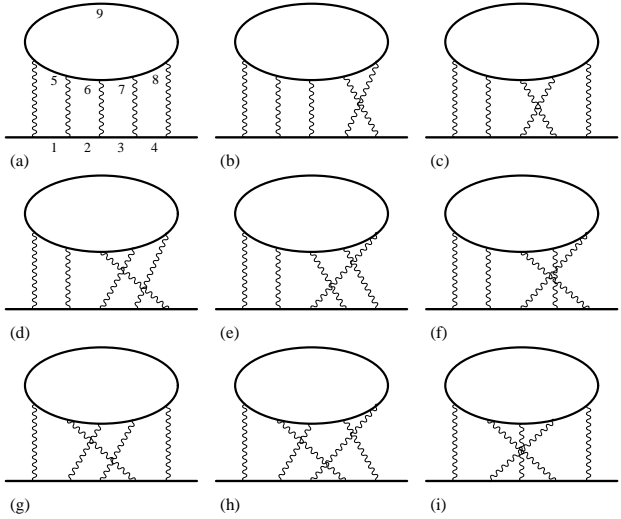}
\caption{\label{X6kall} Subset VI[k] }
\end{figure}

Another point of interest is that its leading logarithmic term
is known (\ref{VIkleading})
and was used to estimate its contribution to
$a_\mu^{(10)}$.
However, to determine the actual contribution of VI(k)
we must find the nonleading term, too.
To answer this question it is best to evaluate VI(k) explicitly.

It turned out that this is not difficult.
Our first step is to reduce the number of independent integrals
to 12 using the Ward-Takahashi identity, and reduce it further
to 9 using the time-reversal symmetry\footnote{ 
We noticed that $X6k_f$ is actually identical with $X6k_b$
while checking the proof. Thus the number of
independent integrals of set VI is 8, not 9.
Fortunately this does not affect our numerical result
given in Eq. (56).}.
Each integral generated by FORM has more than 90,000 terms
occupying about 30,000 lines of FORTRAN code.
This is certainly huge, but not unmanageable since it is
only 30 times larger than typical eighth-order integrals.
We also note that this subset is particularly simple in the sense
that it is entirely free from UV- and IR-divergences.
Numerical integration over 13-dimensional Feynman parameter
space can be handled without difficulty by VEGAS.
The result of numerical integration is listed in Table \ref{table6k},
from which we obtain
\begin{equation}
A_2 [VI(k)] = 97.123~(62).
\label{A2VIk}
\end{equation}
Clearly the previous estimate was an overestimate by about 100.

Another possibly large term is VI(j) [162 vertex diagrams]
which we decided to evaluate explicitly. See Fig. 7.
With the help of Ward-Takahashi identity and time-reversal
invariance it can be represented by four independent integrals.
FORM generated about 42,000 terms 
for each integral occupying about 18,000 lines
of FORTRAN code.
The result of numerical integration is listed in Table \ref{table6j},
from which we obtain
\begin{equation}
A_2 [VI(j)] = -25.505~(20).
\label{A2VIj}
\end{equation}

The subsets yet to be evaluated are VI(d), VI(g), and VI(h).
We foresee no technical problem in dealing with these subsets.

Some results described in this section can be compared with
the results obtained by the renormalization group method.
See \cite{kataev2} for details.

%%%%%%%%%%%%%%%%%%%%%%%%%%%%%%%%%%%%%%%%%%%%%%%%%%%%%%%%
\section{Contribution to electron g-2  }
\label{sec:electron}
%%%%%%%%%%%%%%%%%%%%%%%%%%%%%%%%%%%%%%%%%%%%%%%%%%%%%%%%

All Tables also contain values of mass-independent contributions
from 958 vertex diagrams belonging to 17 gauge-invariant subsets.
These are actually contributions to $A_1^{(10)}$, namely the electron g-2.
Their values, including residual renormalization terms
whenever they are required, are listed below:
\begin{eqnarray}
A_1 [I(a)] &=& 4.7094 ~(6) \times 10^{-4},    
\nonumber   \\
A_1 [I(b)] &=& 7.0108 ~(7) \times 10^{-3},    
\nonumber   \\
A_1 [I(c)] &=& 2.3468 ~(2) \times 10^{-2},    
\nonumber   \\
A_1 [I(d)] &=& 4.4517 ~(5) \times 10^{-3},    
\nonumber   \\
A_1 [I(e)] &=& 1.0296 ~(4) \times 10^{-2},    
\nonumber   \\
A_1 [I(f)] &=& 8.4459 ~(14) \times 10^{-3}.    
\label{EsetI}
\end{eqnarray}
\begin{eqnarray}
A_1 [II(a)] &=& 4.130 ~(90) \times 10^{-3},    
\nonumber   \\
A_1 [II(b)] &=&-5.422 ~(4) \times 10^{-2},    
\nonumber   \\
A_1 [II(f)] &=&-2.434 ~(2).
\label{EsetII}
\end{eqnarray}
\begin{eqnarray}
A_1 [VI(a)] &=& 1.0417 ~(4) ,    
\nonumber   \\
A_1 [VI(b)] &=& 1.3473 ~(3) ,    
\nonumber   \\
A_1 [VI(c)] &=&-2.5922 ~(34) ,    
\nonumber   \\
A_1 [VI(e)] &=&-0.4312 ~(6) ,    
\nonumber   \\
A_1 [VI(f)] &=& 0.7703 ~(24) ,    
\nonumber   \\
A_1 [VI(i)] &=&-0.0438 ~(11) ,    
\nonumber   \\
A_1 [VI(j)] &=&-0.2288 ~(17) ,    
\nonumber   \\
A_1 [VI(k)] &=& 0.6802 ~(38) .    
\label{EsetVI}
\end{eqnarray}

%%%%%%%%%%%%%%%%%%%%%%%%%%%%%%%%%%%%%%%%%%%%%%%%%%%%%%%%
\section{Discussion  }
\label{sec:feasible}
%%%%%%%%%%%%%%%%%%%%%%%%%%%%%%%%%%%%%%%%%%%%%%%%%%%%%%%%

Let us first add up all 
terms contributing to $A_2^{(10)} (m_\mu/m_e )$
of muon g-2 evaluated by numerical integration.  From
(\ref{A2Ia}), (\ref{A2Ib-c}), (\ref{A2Ie}), (\ref{A2Id}), 
(\ref{A2If}), (\ref{A2IIa}), (\ref{A2IIb}), (\ref{A2IIf}), 
(\ref{A2VIa}), (\ref{A2VIb}),
(\ref{6c}), (\ref{6e}), (\ref{6f}), (\ref{6i}), 
(\ref{A2VIk}), (\ref{A2VIj}),
which represent the
contributions of 2958 Feynman diagrams belonging to 17
gauge-invariant sets, we obtain
\begin{equation}
A_2^{(10)} (m_\mu/m_e) [partial~sum]  = 662.50~(27).
\label{partialsum}
\end{equation}
The uncertainties from these diagrams,
which include all dominant sources of uncertainties considered
previously, have been reduced to an insignificant level.
Note, however, that some terms which were not included
in previous estimates turned out to be not negligible.

Of course, the real value of 
$A_2^{(10)} (m_\mu/m_e)$ is not known until remaining diagrams
are evaluated.
However, they
have no known mechanism for giving rise to large values and
likely to remain modest in size and uncertainty.
We therefore expect that the final value will stay within the range
\begin{equation}
A_2^{(10)} (m_\mu/m_e) [estimate]  = 663~(20).
\label{a2estimate}
\end{equation}
This will reduce (\ref{newQEDvalue}) by $1.81 \times 10^{-11}$. 

Our next step is to evaluate  6122 vertex diagrams from
the remaining 14 gauge-invariant subsets. 
Many of these diagrams can be integrated by means of
available information on $\Pi$ and $\Lambda$.
They include:

\begin{enumerate}
\item  Subsets I(f), I(g), I(h) of Set I.
\item  Subsets II(c), II(d) of Set II.
\item  Subsets III(a) and III(b) of Set III.
\item  All diagrams of Set IV,
which can be evaluated by simple modification
of codes of Group V of eighth-order diagrams.
\end{enumerate}

\noindent
These subsets are on our next time schedule.

The remainder of diagrams are more difficult to evaluate
for the following reasons:

\begin{enumerate}

\item  Subsets I(i) and I(j) require the knowledge of the eighth-order
v-p spectral function  $\Pi_8$ which have not yet been constructed.
\item  Diagrams of set II(e) require construction of radiatively-corrected
light-by-light scattering subdiagrams, which are not yet available.
\item  Subset III[c] contains $\Lambda_4$ internally.
\item  Subsets VI(d), VI(g), and VI(h)
have no lower-order structure upon which they
can be built.
\end{enumerate}

\noindent
We foresee no intractable barrier for their evaluation.

As far as the electron g-2 is concerned what is most important 
is the Set V which have no lower-order fermion loop structure.
The integrands of this set are gigantic and require an enormous
number of UV and IR subtraction terms.
Our experience with the subsets VI(k) and VI(j) indicates, however,
that their sizes are still manageable with available computers.
What is really crucial for their evaluation, however, 
is that all steps of construction of
integrand must be fully automated.
Thus far we have succeeded in obtaining a code that 
handles renormalization of ultraviolet divergence automatically \cite{ahkn}.
For the moment the infrared divergence is treated by giving
a small cutoff mass $\lambda$ to the photon.
A code for cutoff-independent treatment of IR divergence
is being developed.

\begin{acknowledgments}

T. K.'s work is supported by the U. S. National Science Foundation under Grant
No. PHY-0098631.  
T. K. thanks the Eminent Scientist Invitation Program
of RIKEN, Japan, for the hospitality extended to him where a part
of this work was carried out.
T. K. is also supported during his stay in Japan by Ministry of Education, 
Science and Culture of Japan,
Grant-in-Aid for Scientific Research on Priority Areas, 13134101.

M. N.'s work is partly supported by 
Japan Society for the Promotion of Science,
%Ministry of Education, Science and Culture of Japan,
Grant-in-Aid for Scientific Research (C) 15540303, 2003-2005.

The numerical work has been carried out on the RIKEN
Super Combined Cluster System (RSCC).

\end{acknowledgments}

%%%%%%%%%%%%%%%%%%%%%%%%%%%%%%%%%%%%%%%%%%%%%%%%%%%%%%%%%%%
%                                                         %
%                    REFERENCES                           %
%                                                         %
%%%%%%%%%%%%%%%%%%%%%%%%%%%%%%%%%%%%%%%%%%%%%%%%%%%%%%%%%%%

\renewcommand{\arraystretch}{0.80}
\begin{table}
\caption{Numerical evaluation of diagrams of subsets
(a), (b), (c), and (e) of Set I.
The notation follows that of \cite{oldtk1} 
with some modification and adaptation. 
$n_F$ is the number of Feynman diagrams
represented by the integral.
 \\
\label{table1}
}
\begin{tabular}{lcrrr}
\hline
\hline
~Integral~~&~~~ $n_F$~~~ &~ ~Value (error)~~~&~Sampling~per~& ~~~No. of~~~ \\ [.1cm]  
&  &~including $n_F$~ ~&~iteration~~~~~& ~~iterations~  \\ [.1cm]   \hline
\\
$M_{2,p2:4}^{(e,e,e,e)}$ &1&$~20.14293~(23)$&~~$1 \times 10^{8}$ \hspace{4mm}~~&180\hspace{4mm}~ \\
$M_{2,p2:4}^{(e,e,e,m)}$&4&~~2.20327~(~9)&~~$1 \times 10^{7}$ \hspace{4mm}~~&80\hspace{4mm}~ \\
$M_{2,p2:4}^{(e,e,m,m)}$&6&~~0.20697~(~2)&~~$1 \times 10^{7}$ \hspace{4mm}~~&20\hspace{4mm}~ \\
$M_{2,p2:4}^{(e,m,m,m)}$&4&~~0.01388~(~1)&~~$1 \times 10^{7}$ \hspace{4mm}~~&20\hspace{4mm}~ \\
$M_{2,p2:4}^{(m,m,m,m)}$&1&~~4.7094~(6)$\times 10^{-4}$&~~$1 \times 10^{7}$ \hspace{4mm}~~&20\hspace{4mm}~ \\
\\
$M_{2,p4,p2:2}^{(e,e,e)}$ &9&$~27.69038~(30)$&~~$4 \times 10^{8}$ \hspace{4mm}~~&190\hspace{4mm}~ \\
$M_{2,p4,p2:2}^{(e,m,e)}$ &18&$~~1.16628~(~9)$&~~$1 \times 10^{7}$ \hspace{4mm}~~&100\hspace{4mm}~ \\
$M_{2,p4,p2:2}^{(e,m,m)}$ &9&$~~0.03182~(~3)$&~~$1 \times 10^{6}$ \hspace{4mm}~~&20\hspace{4mm}~ \\
$M_{2,p4,p2:2}^{(m,e,e)}$ &9&$~~1.61436~(~6)$&~~$1 \times 10^{7}$ \hspace{4mm}~~&120\hspace{4mm}~ \\
$M_{2,p4,p2:2}^{(m,m,e)}$ &18&$~~0.16470~(~5)$&~~$1 \times 10^{6}$ \hspace{4mm}~~&20\hspace{4mm}~ \\
$M_{2,p4,p2:2}^{(m,m,m)}$ &9&~~7.0108~(7)$\times 10^{-3}$&~~$1 \times 10^{7}$ \hspace{4mm}~~&20\hspace{4mm}~ \\
\\
$M_{2,p4:2}^{(e,e)}$ &9&$~~4.74212~(14)$&~~$1 \times 10^{8}$ \hspace{4mm}~~&220\hspace{4mm}~ \\
$M_{2,p4:2}^{(e,m)}$ &18&$~~0.39926~(~3)$&~~$1 \times 10^{7}$ \hspace{4mm}~~&120\hspace{4mm}~ \\
$M_{2,p4:2}^{(m,m)}$ &9&~~2.3468~(2)$\times 10^{-2}$&~~$1 \times 10^{7}$ \hspace{4mm}~~&20\hspace{4mm}~ \\
\\
$M_{2,p6p2}^{(e,e)}$ &30&$~-1.20841~(70)$&~~$1 \times 10^{8}$ \hspace{4mm}~~&100\hspace{4mm}~ \\
$M_{2,p6p2}^{(e,m)}$ &30&$~-0.02110~(~4)$&~~$1 \times 10^{7}$ \hspace{4mm}~~&120\hspace{4mm}~ \\
$M_{2,p6p2}^{(m,e)}$ &30&$~~0.01031~(~1)$&~~$1 \times 10^{6}$ \hspace{4mm}~~&20\hspace{4mm}~ \\
$M_{2,p6p2}^{(m,m)}$ &30&~~1.0296~(4)$\times 10^{-2}$&~~$1 \times 10^{7}$ \hspace{4mm}~~&20\hspace{4mm}~ \\
\\
\hline
\hline
\end{tabular}
\end{table}
\renewcommand{\arraystretch}{1}

\renewcommand{\arraystretch}{0.80}
\begin{table}
\caption{Numerical evaluation of diagrams of subsets (d) of Set I.
The notation follows that of \cite{oldtk1} 
with some modification and adaptation. 
$n_F$ is the number of Feynman diagrams
represented by the integral.
Subscripts $p4A$, $p4B$ refer to part of fourth-order $v-p$ $\Pi_4$
containing vertex correction and self-energy insertion, respectively.
The last four lines are values obtained using the exact sixth-order
spectral function $\Pi_{4(2)}$.
 \\
\label{table1d}
}
\begin{tabular}{lcrrr}
\hline
\hline
~Integral~~&~~~ $n_F$~~~ &~ ~Value (error)~~~&~Sampling~per~& ~~~No. of~~~ \\ [.1cm]  
&  &~including $n_F$~ ~&~iteration~~~~~& ~~iterations~  \\ [.1cm]   \hline
\\
$M_{2,p4A(p2)p2}^{(e(e),e)}$ &2&$~~2.63064~(72)$&~~$1 \times 10^{7}$ \hspace{4mm}~~&80\hspace{4mm}~ \\
$M_{2,p4A(p2)p2}^{(m(e),e)}$ &2&$~~0.76997~(~7)$&~~$1 \times 10^{7}$ \hspace{4mm}~~&40\hspace{4mm}~ \\
$M_{2,p4A(p2)p2}^{(e(m),e)}$ &2&$~~0.12703~(~2)$&~~$1 \times 10^{7}$ \hspace{4mm}~~&40\hspace{4mm}~ \\
$M_{2,p4A(p2)p2}^{(m(m),e)}$ &2&$~~7.3352~(~7)\times 10^{-2}$&~~$1 \times 10^{7}$ \hspace{4mm}~~&40\hspace{4mm}~ \\
$M_{2,p4A(p2)p2}^{(e(e),m)}$ &2&$~~1.3211~(18)\times 10^{-2}$&~~$1 \times 10^{7}$ \hspace{4mm}~~&79\hspace{4mm}~ \\
$M_{2,p4A(p2)p2}^{(m(e),m)}$ &2&$~~3.5428~(~4)\times 10^{-2}$&~~$1 \times 10^{7}$ \hspace{4mm}~~&40\hspace{4mm}~ \\
$M_{2,p4A(p2)p2}^{(e(m),m)}$ &2&$~~5.0276~(~6)\times 10^{-3}$&~~$1 \times 10^{7}$ \hspace{4mm}~~&40\hspace{4mm}~ \\
$M_{2,p4A(p2)p2}^{(m(m),m)}$ &2&$~~3.8330~(~4)\times 10^{-3}$&~~$1 \times 10^{7}$ \hspace{4mm}~~&40\hspace{4mm}~ \\
\\
$M_{2,p4B(p2)p2}^{(e(e),e)}$ &4&$~~5.51053~(70)$&~~$1 \times 10^{7}$ \hspace{4mm}~~&80\hspace{4mm}~ \\
$M_{2,p4B(p2)p2}^{(m(e),e)}$ &4&$~~0.63490~(~9)$&~~$1 \times 10^{7}$ \hspace{4mm}~~&40\hspace{4mm}~ \\
$M_{2,p4B(p2)p2}^{(e(m),e)}$ &4&$~~3.9114~(47) \times 10^{-3}$&~~$1 \times 10^{7}$ \hspace{4mm}~~&29\hspace{4mm}~ \\
$M_{2,p4B(p2)p2}^{(m(m),e)}$ &4&$~~1.1129~(~4) \times 10^{-2}$&~~$1 \times 10^{7}$ \hspace{4mm}~~&29\hspace{4mm}~ \\
$M_{2,p4B(p2)p2}^{(e(e),m)}$ &4&$~~0.15798~(~2)$&~~$1 \times 10^{7}$ \hspace{4mm}~~&80\hspace{4mm}~ \\
$M_{2,p4B(p2)p2}^{(m(e),m)}$ &4&$~~3.2133~(~5)\times 10^{-2}$&~~$1 \times 10^{7}$ \hspace{4mm}~~&40\hspace{4mm}~ \\
$M_{2,p4B(p2)p2}^{(e(m),m)}$ &4&$~~2.6449~(19) \times 10^{-4}$&~~$1 \times 10^{7}$ \hspace{4mm}~~&40\hspace{4mm}~ \\
$M_{2,p4B(p2)p2}^{(m(m),m)}$ &4&$~~6.1873~(17) \times 10^{-4}$&~~$1 \times 10^{7}$ \hspace{4mm}~~&40\hspace{4mm}~ \\
\\
\\
$M_{2,p4(p2)p2}^{(e(e),e)}$ &6&$~~7.45270~(88)$&~~$1 \times 10^{8}$ \hspace{4mm}~~&140\hspace{4mm}~ \\
$M_{2,p4(p2)p2}^{(e(e),m)}$ &6&$~~0.15853~(10)$&~~$1 \times 10^{7}$ \hspace{4mm}~~&120\hspace{4mm}~ \\
$M_{2,p4(p2)p2}^{(m(m),e)}$ &6&$~~0.07173~(~3)$&~~$1 \times 10^{6}$ \hspace{4mm}~~&20\hspace{4mm}~ \\
$M_{2,p4(p2)p2}^{(m(m),m)}$ &6&~~3.8028~(5)$\times 10^{-3}$&~~$1 \times 10^{7}$ \hspace{4mm}~~&20\hspace{4mm}~ \\
\\
\hline
\hline
\end{tabular}
\end{table}
\renewcommand{\arraystretch}{1}

\renewcommand{\arraystretch}{0.80}
\begin{table}
\caption{Numerical evaluation of diagrams of subsets (f) of Set I.
The notation follows that of \cite{oldtk1} 
with some modification and adaptation. 
$n_F$ is the number of Feynman diagrams
represented by the integral.
Subscripts $p4A$, $p4B$ refer to part of fourth-order $v-p$ $\Pi_4$
containing vertex correction and self-energy insertion, respectively.
 \\
\label{table1f}
}
\begin{tabular}{lcrrr}
\hline
\hline
~Integral~~&~~~ $n_F$~~~ &~ ~Value (error)~~~&~Sampling~per~& ~~~No. of~~~ \\ [.1cm]  
&  &~including $n_F$~ ~&~iteration~~~~~& ~~iterations~  \\ [.1cm]   \hline
\\
$\Delta M_{2,p4A(p2:2)}^{(e(e,e))}$ &1&$~~1.99747~(10)$&~~$4 \times 10^{7}$ \hspace{4mm}~~&240\hspace{4mm}~ \\
$\Delta M_{2,p4A(p2:2)}^{(e(e,m))}$ &2&$~~0.15863~(3)$&~~$1 \times 10^{7}$ \hspace{4mm}~~&20\hspace{4mm}~ \\
$\Delta M_{2,p4A(p2:2)}^{(e(m,m))}$ &1&$~~1.0612~(2) \times 10^{-2}$&~~$1 \times 10^{7}$ \hspace{4mm}~~&20\hspace{4mm}~ \\
$\Delta M_{2,p4A(p2:2)}^{(m(e,e))}$ &1&$~~0.44361~(6)$&~~$1 \times 10^{7}$ \hspace{4mm}~~&20\hspace{4mm}~ \\
$\Delta M_{2,p4A(p2:2)}^{(m(e,m))}$ &2&$~~9.4542~(13) \times 10^{-2}$&~~$1 \times 10^{7}$ \hspace{4mm}~~&20\hspace{4mm}~ \\
$\Delta M_{2,p4A(p2:2)}^{(m(m,m))}$ &1&$~~8.1763~(13) \times 10^{-3}$&~~$1 \times 10^{7}$ \hspace{4mm}~~&20\hspace{4mm}~ \\
\\
$\Delta M_{2,p4B(p2:2)}^{(e(e,e))}$ &2&$~~0.94959~(11)$&~~$1 \times 10^{7}$ \hspace{4mm}~~&340\hspace{4mm}~ \\
$\Delta M_{2,p4B(p2:2)}^{(e(e,m))}$ &4&$~~2.5573~(56) \times 10^{-3}$&~~$1 \times 10^{7}$ \hspace{4mm}~~&20\hspace{4mm}~ \\
$\Delta M_{2,p4B(p2:2)}^{(e(m,m))}$ &2&$~~1.6448~(275) \times 10^{-5}$&~~$1 \times 10^{7}$ \hspace{4mm}~~&20\hspace{4mm}~ \\
$\Delta M_{2,p4B(p2:2)}^{(m(e,e))}$ &2&$~~0.26023~(6)$&~~$1 \times 10^{7}$ \hspace{4mm}~~&20\hspace{4mm}~ \\
$\Delta M_{2,p4B(p2:2)}^{(m(e,m))}$ &4&$~~1.1060~(5) \times 10^{-2}$&~~$1 \times 10^{7}$ \hspace{4mm}~~&20\hspace{4mm}~ \\
$\Delta M_{2,p4B(p2:2)}^{(m(m,m))}$ &2&$~~2.6957~(32) \times 10^{-4}$&~~$1 \times 10^{7}$ \hspace{4mm}~~&20\hspace{4mm}~ \\
\\
\hline
\hline
\end{tabular}
\end{table}
\renewcommand{\arraystretch}{1}

\renewcommand{\arraystretch}{0.9}
\begin{table}
\caption{Numerical evaluation of set II(a) which consists of II($a_1$)
and II($a_2$).
The set II($a_1$) has contributions denoted by $\Delta_1$ and
set II($a_2$) has contributions denoted by $\Delta_2$.
The symbol ()() indicates photon lines in 
which v-p loops $\Pi$ are inserted. 
$n_F$ is the number of Feynman diagrams
represented by the integral.
t.r. is time-reversed Feynman diagram.
\\
\label{table2a}
}
\begin{tabular}{lcrrr}
\hline
\hline
~Integral~~&~~~ $n_F$~~~ &~ ~Value (error)~~~&~Sampling~per~& ~~~No. of~~~ \\ [.1cm]  
&  &~including~$n_F$~~~&~iteration~~~~~& ~~iterations~  \\ [.1cm]   \hline
\\
$\Delta_1 M_{4a,p2:3}^{(e,e,e)()}$ + t.r.&6&$15.15611~(283)$&~~$1 \times 10^{8}$ \hspace{4mm}~~&160\hspace{4mm}~ \\
$\Delta_1 M_{4a,p2:3}^{(e,e,m)()}$ + t.r.&18&$1.31796~(~49)$&~~$1 \times 10^{8}$ \hspace{4mm}~~&200\hspace{4mm}~ \\
$\Delta_1 M_{4a,p2:3}^{(e,m,m)()}$ + t.r.&18&$0.09109~(~13)$&~~$1 \times 10^{8}$ \hspace{4mm}~~&100\hspace{4mm}~ \\
$\Delta_1 M_{4a,p2:3}^{(m,m,m)()}$ + t.r.&6&$0.00413~(~~9)$&~~$1 \times 10^{7}$ \hspace{4mm}~~&20\hspace{4mm}~ \\
\\
$\Delta_1 M_{4b,p2:3}^{(e,e,e)()}$ +$\Delta_1 M_{4b,p2:3}^{()(e,e,e)}$
&6&$-30.09130~(191)$&~~$4 \times 10^{7}$ \hspace{4mm}~~&200\hspace{4mm}~ \\
$\Delta_1 M_{4b,p2:3}^{(e,e,m)()}$ +$\Delta_1 M_{4b,p2:3}^{()(e,e,m)}$
&18&$-6.38188~(~26)$&~~$1 \times 10^{8}$ \hspace{4mm}~~&100\hspace{4mm}~ \\
$\Delta_1 M_{4b,p2:3}^{(e,m,m)()}$ +$\Delta_1 M_{4b,p2:3}^{()(e,m,m)}$
&18&$-0.84226~(~~4)$&~~$1 \times 10^{8}$ \hspace{4mm}~~&120\hspace{4mm}~ \\
$\Delta_1 M_{4b,p2:3}^{(m,m,m)()}$ +$\Delta_1 M_{4b,p2:3}^{()(m,m,m)}$
&6&$-0.05422~(~~4)$&~~$1 \times 10^{7}$ \hspace{4mm}~~&20\hspace{4mm}~ \\
\\
$\Delta_2 M_{4a,p2:2,p2}^{(e,e)(e)}$ + t.r.&6&$12.23457~(99)$&~~$2 \times 10^{8}$ \hspace{4mm}~~&370\hspace{4mm}~ \\
$\Delta_2 M_{4a,p2:2,p2}^{(e,e)(m)}$ + t.r.&6&$0.01971~(34)$&~~$1 \times 10^{7}$ \hspace{4mm}~~&120\hspace{4mm}~ \\
$\Delta_2 M_{4a,p2:2,p2}^{(e,m)(e)}$ + t.r.&12&$0.27276(35)$&~~$2 \times 10^{7}$ \hspace{4mm}~~&380\hspace{4mm}~ \\
$\Delta_2 M_{4a,p2:2,p2}^{(m,m)(e)}$ + t.r.&6&$-0.01408(27)$&~~$1 \times 10^{7}$ \hspace{4mm}~~&20\hspace{4mm}~ \\
$\Delta_2 M_{4a,p2:2,p2}^{(e,m)(m)}$ + t.r.&12&$-0.08222(22)$&~~$1 \times 10^{7}$ \hspace{4mm}~~&20\hspace{4mm}~ \\
$\Delta_2 M_{4a,p2:2,p2}^{(m,m)(m)}$ + t.r.&6&$-0.00998(~3)$&~~$1 \times 10^{7}$ \hspace{4mm}~~&20\hspace{4mm}~ \\
\\
$\Delta_2 M_{4b,p2:2,p2}^{(e,e)(e)}$ +$\Delta_2 M_{4b,p2,p2:2}^{(e)(e,e)}$
&6&$-28.69942~(~79)$&~~$1 \times 10^{8}$ \hspace{4mm}~~&380\hspace{4mm}~ \\
$\Delta_2 M_{4b,p2:2,p2}^{(e,e)(m)}$ +$\Delta_2 M_{4b,p2,p2:2}^{(m)(e,e)}$
&6&$-1.72076~(14)$&~~$1 \times 10^{7}$ \hspace{4mm}~~&100\hspace{4mm}~ \\
$\Delta_2 M_{4b,p2:2,p2}^{(e,m)(e)}$ +$\Delta_2 M_{4b,p2,p2:2}^{(e)(e,m)}$
&12&$-3.71561~(32)$&~~$1 \times 10^{7}$ \hspace{4mm}~~&120\hspace{4mm}~ \\
$\Delta_2 M_{4b,p2:2,p2}^{(m,m)(e)}$ +$\Delta_2 M_{4b,p2,p2:2}^{(e)(m,m)}$
&6&$-0.23956~(~8)$&~~$1 \times 10^{7}$ \hspace{4mm}~~&20\hspace{4mm}~ \\
$\Delta_2 M_{4b,p2:2,p2}^{(e,m)(m)}$ +$\Delta_2 M_{4b,p2,p2:2}^{(m)(e,m)}$
&12&$-0.37976~(~7)$&~~$1 \times 10^{7}$ \hspace{4mm}~~&20\hspace{4mm}~ \\
$\Delta_2 M_{4b,p2:2,p2}^{(m,m)(m)}$ +$\Delta_2 M_{4b,p2,p2:2}^{(m)(m,m)}$
&6&$-0.03619~(~1)$&~~$1 \times 10^{7}$ \hspace{4mm}~~&20\hspace{4mm}~ \\
\\
\hline
\hline
\end{tabular}
\end{table}

\renewcommand{\arraystretch}{1.0}
\begin{table}
\caption{Numerical evaluation of diagrams of set II(b).
$\Delta_1$ indicates contributions in which $\Pi$'s act on the same
photon line while $\Delta_2$ refers to those in which $\Pi$'s
act on different photon lines.
The symbol ()() indicates photon lines in 
which v-p loops $\Pi$ are inserted. 
$n_F$ is the number of Feynman diagrams
represented by the integral.
t.r. is time-reversed Feynman diagram.
\\
\label{table2b}
}
\begin{tabular}{lcrrr}
\hline
\hline
~Integral~~&~~~ $n_F$~~~ &~ ~Value (error)~~~&~Sampling~per~& ~~~No. of~~~ \\ [.1cm]  
&  &~including~$n_F$~~~&~iteration~~~~~& ~~iterations~  \\ [.1cm]   \hline
\\
$\Delta_1 M_{4a,p4p2}^{(e,e)()}$+($p4 \leftrightarrow p2$)+t.r.&36&$12.65000~(135)$&~~$4 \times 10^{8}$ \hspace{4mm}~~&220\hspace{4mm}~ \\
$\Delta_1 M_{4a,p4p2}^{(e,m)()}$+($p4 \leftrightarrow p2$)+t.r.&36&$0.30749~(~22)$&~~$4 \times 10^{7}$ \hspace{4mm}~~&140\hspace{4mm}~ \\
$\Delta_1 M_{4a,p4p2}^{(m,e)()}$+($p4 \leftrightarrow p2$)+t.r.&36&$0.84809~(~23)$&~~$1 \times 10^{8}$ \hspace{4mm}~~&160\hspace{4mm}~ \\
$\Delta_1 M_{4a,p4p2}^{(m,m)()}$+($p4 \leftrightarrow p2$)+t.r.&36&$0.04424~(~33)$&~~$1 \times 10^{7}$ \hspace{4mm}~~&20\hspace{4mm}~ \\
\\
$\Delta_1 M_{4b,p4p2}^{(e,e)()}$ +$\Delta_1 M_{4b,p4p2}^{()(e,e)}$
+($p4 \leftrightarrow p2$)
&36&$-19.70781~(143)$&~~$1 \times 10^{8}$ \hspace{4mm}~~&220\hspace{4mm}~ \\
$\Delta_1 M_{4b,p4p2}^{(e,m)()}$ +$\Delta_1 M_{4b,p4p2}^{()(e,m)}$
+($p4 \leftrightarrow p2$)
&36&$-1.23984~(~10)$&~~$4 \times 10^{7}$ \hspace{4mm}~~&140\hspace{4mm}~ \\
$\Delta_1 M_{4b,p4p2}^{(m,e)()}$ +$\Delta_1 M_{4b,p4p2}^{()(m,e)}$
+($p4 \leftrightarrow p2$)
&36&$-3.06974~(~~9)$&~~$1 \times 10^{8}$ \hspace{4mm}~~&200\hspace{4mm}~ \\
$\Delta_1 M_{4b,p4p2}^{(m,m)()}$ +$\Delta_1 M_{4b,p4p2}^{()(m,m)}$
+($p4 \leftrightarrow p2$)
&36&$-0.30984~(~15)$&~~$1 \times 10^{7}$ \hspace{4mm}~~&20\hspace{4mm}~ \\
\\
$\Delta_2 M_{4a,p4p2}^{(e)(e)}$+t.r.&18&$5.89471~(107)$&~~$1 \times 10^8$ \hspace{4mm}~~&240\hspace{4mm}~ \\
$\Delta_2 M_{4a,p4p2}^{(e)(m)}$+t.r.&18&$0.09286~(~16)$&~~$1 \times 10^7$ \hspace{4mm}~~&140\hspace{4mm}~ \\
$\Delta_2 M_{4a,p4p2}^{(m)(e)}$+t.r.&18&$0.21886~(~23)$&~~$1 \times 10^7$ \hspace{4mm}~~&180\hspace{4mm}~ \\
$\Delta_2 M_{4a,p4p2}^{(m)(m)}$+t.r.&18&$-0.01723~(~~5)$&~~$1 \times 10^7$ \hspace{4mm}~~&20\hspace{4mm}~ \\
\\
$\Delta_2 M_{4b,p4p2}^{(e)(e)}$ +($p4 \leftrightarrow p2$)
&18&$-9.83570~(97)$&~~$4 \times 10^7$ \hspace{4mm}~~&260\hspace{4mm}~ \\
$\Delta_2 M_{4b,p4p2}^{(e)(m)}$ +($p4 \leftrightarrow p2$)
&18&$-0.56867~(~8)$&~~$1 \times 10^7$ \hspace{4mm}~~&120\hspace{4mm}~ \\
$\Delta_2 M_{4b,p4p2}^{(m)(e)}$ +($p4 \leftrightarrow p2$)
&18&$-1.36297~(13)$&~~$1 \times 10^7$ \hspace{4mm}~~&100\hspace{4mm}~ \\
$\Delta_2 M_{4b,p4p2}^{(m)(m)}$ +($p4 \leftrightarrow p2$)
&18&$-0.10756~(~2)$&~~$1 \times 10^7$ \hspace{4mm}~~&20\hspace{4mm}~ \\
\\
\hline
\hline
\end{tabular}
\end{table}
\renewcommand{\arraystretch}{1}

\renewcommand{\arraystretch}{1.0}
\begin{table}
\caption{ Numerical evaluation of diagrams for auxiliary quantities
needed to evaluate contribution of Set I and Set II. \\
\label{table3}
}
\begin{tabular}{lcrrr}
\hline
\hline
~Integral~~&~~~ $n_F$~~~ &~ ~Value (error)~~~&~Sampling~per~& ~~~No. of~~~ \\ [.1cm]  
&  &~including $n_F$~ ~&~iteration~~~~~& ~~iterations~  \\ [.1cm]   \hline
\\
$\Delta B_{2,2:3}^{(e,e,e)}$&1&$16.15765~(75)$&~~$1 \times 10^{7}$ \hspace{4mm}~~&100\hspace{4mm}~ \\
$\Delta B_{2,2:3}^{(e,e,m)}$&3&$2.71594~(12)$&~~$1 \times 10^{7}$ \hspace{4mm}~~&100\hspace{4mm}~ \\
$\Delta B_{2,2:3}^{(e,m,m)}$&3&$0.36136~(~4)$&~~$1 \times 10^{7}$ \hspace{4mm}~~&40\hspace{4mm}~ \\
$\Delta B_{2,2:3}^{(m,m,m)}$&1&$0.02381~(~1)$&~~$1 \times 10^{7}$ \hspace{4mm}~~&20\hspace{4mm}~ \\
\\
$\Delta B_{2,p4p2}^{(e,e)}$&6&$13.27621~(93)$&~~$1 \times 10^{7}$ \hspace{4mm}~~&100\hspace{4mm}~ \\
$\Delta B_{2,p4p2}^{(e,m)}$&6&$0.53279~(~5)$&~~$1 \times 10^{7}$ \hspace{4mm}~~&80\hspace{4mm}~ \\
$\Delta B_{2,p4p2}^{(m,e)}$&6&$1.31884~(~5)$&~~$1 \times 10^{7}$ \hspace{4mm}~~&120\hspace{4mm}~ \\
$\Delta B_{2,p4p2}^{(m,m)}$&6&$0.13066~(~2)$&~~$1 \times 10^{7}$ \hspace{4mm}~~&20\hspace{4mm}~ \\
\\
$\Delta B_{2,p2:2}^{(e(e,e))}$&1&$0.02791~(~1)$&~~$1 \times 10^{6}$ \hspace{4mm}~~&40\hspace{4mm}~ \\
$\Delta B_{2,p2:2}^{(m(e,e))}$&1&$5.33035~(20)$&~~$1 \times 10^{7}$ \hspace{4mm}~~&100\hspace{4mm}~ \\
$\Delta B_{2,p2:2}^{(m(m,e))}$&2&$0.47208~(~9)$&~~$1 \times 10^{6}$ \hspace{4mm}~~&40\hspace{4mm}~ \\
$\Delta B_{2,p2:2}^{(e(m,e))}$&2&$3.6159~(8) \times 10^{-5}$&~~$1 \times 10^{6}$ \hspace{4mm}~~&40\hspace{4mm}~ \\
$\Delta B_{2,p2:2}^{(e(m,m))}$&1&$8.3020~(36) \times 10^{-7}$&~~$1 \times 10^{6}$ \hspace{4mm}~~&20\hspace{4mm}~ \\
\\
\hline
\hline
\end{tabular}
\end{table}
\renewcommand{\arraystretch}{1}

\renewcommand{\arraystretch}{0.9}
\begin{table}
\caption{Numerical evaluation of diagrams of set II(f) in Version A.
The suffix $p$ indicates insertion of $\Pi_2$ in the photon lines
connecting the muon line and the $l-l$ loop.
$A_2 [2f]^{(x,y)} \equiv 
\Delta M_{8LLJp}^{(x,y)} +
\Delta M_{8LLKp}^{(x,y)} +
\Delta M_{8LLLp}^{(x,y)} $.
$r = \Delta M_{8LLxp}/\Delta M_{8LLx}$, where $x = J, K, L$,
in column 6 is for comparison with the enhancement factor
$4K_\eta$ for set II(f) discussed in Sec. \ref{sec:leading}.
The logarithmic enhancement comes from the $v-p$ loop only.
This is consistent with $r \simeq 11 \sim 15$ for (e,e), (me) and
$r \simeq 0.7 \sim 2$ for (e,m), (m,m).\\
\label{table2f}
}
\begin{tabular}{lcrrrr}
\hline
\hline
~Integral~~& ~~~$n_F$~~~ &~ ~Value (error)~~~&~Sampling~per~& ~~~No. of~~~\hspace{4mm} &~$r$~ \\ [.1cm]  
&  &~including~$n_F$~~~&~iteration~~~~~& ~~iterations~ \hspace{4mm} &  \\ [.1cm]   \hline
\\
$\Delta M_{8LLJp}^{(e,e)}$&24&$70.6567~(254)$&~~$4 \times 10^{7}$ \hspace{4mm}~~&220\hspace{14mm}~&11.05 \\
$\Delta M_{8LLJp}^{(m,e)}$&24&$35.0760~(~78)$&~~$2 \times 10^{7}$ \hspace{4mm}~~&200\hspace{14mm}~&13.74 \\
$\Delta M_{8LLJp}^{(e,m)}$&24&$6.0056~(~26)$&~~$2 \times 10^{7}$ \hspace{4mm}~~&200\hspace{14mm}~&0.94 \\
$\Delta M_{8LLJp}^{(m,m)}$&24&$3.7717~(~17)$&~~$1 \times 10^{7}$ \hspace{4mm}~~&140\hspace{14mm}~&1.48 \\
\\
$\Delta M_{8LLKp}^{(e,e)}$&24&$-87.8367~(291)$&~~$8 \times 10^{7}$ \hspace{4mm}~~&210\hspace{14mm}~&11.29 \\
$\Delta M_{8LLKp}^{(m,e)}$&24&$-26.1793~(~83)$&~~$2 \times 10^{7}$ \hspace{4mm}~~&220\hspace{14mm}~&13.97 \\
$\Delta M_{8LLKp}^{(e,m)}$&24&$-5.4760~(~29)$&~~$2 \times 10^{7}$ \hspace{4mm}~~&230\hspace{14mm}~&0.70 \\
$\Delta M_{8LLKp}^{(m,m)}$&24&$-2.9311~(~17)$&~~$1 \times 10^{7}$ \hspace{4mm}~~&140\hspace{14mm}~&1.56 \\
\\
$\Delta M_{8LLLp}^{(e,e)}$&24&$-39.9514~(287)$&~~$4 \times 10^{7}$ \hspace{4mm}~~&310\hspace{14mm}~&13.07 \\
$\Delta M_{8LLLp}^{(m,e)}$&24&$-24.5877~(~81)$&~~$2 \times 10^{7}$ \hspace{4mm}~~&220\hspace{14mm}~&14.73 \\
$\Delta M_{8LLLp}^{(e,m)}$&24&$-5.2468~(~29)$&~~$2 \times 10^{7}$ \hspace{4mm}~~&200\hspace{14mm}~&1.71 \\
$\Delta M_{8LLLp}^{(m,m)}$&24&$-3.2771~(~17)$&~~$1 \times 10^{7}$ \hspace{4mm}~~&140\hspace{14mm}~&1.96 \\
\\
\\
$A_2 [2f]^{(e,e)}$&72&-57.1314~(481)&~&~& \\
$A_2 [2f]^{(m,e)}$&72&-15.6910~(140)&~&~& \\
$A_2 [2f]^{(e,m)}$&72&-4.7172~(~49)&~&~& \\
$A_2 [2f]^{(m,m)}$&72&-2.4365~(~29)&~&~& \\
\\
\\
\hline
\hline
\end{tabular}
\end{table}
\renewcommand{\arraystretch}{1}

\renewcommand{\arraystretch}{1.0}
\begin{table}
\caption{Numerical evaluation of diagrams of set II(f) in Version B.
The symbol $p$ indicates insertion of $\Pi_2$ in the photon lines
connecting the muon line and the $light-light$ loop.
$JKLp^{(x,y)} \equiv \Delta M_{8LLJp}^{(x,y)} 
+ \Delta M_{8LLKp}^{(x,y)}
+ \Delta M_{8LLLp}^{(x,y)}$.
The suffix 2 or 13 
below indicates the muon line into which
the magnetic field vertex is inserted. 
$A_2 [2f]^{(x,y)} \equiv 
JKLp_{2}^{(x,y)} +
JKLp_{13}^{(x,y)} $.
$r = JKLp/JKL$
in column 6 is for comparison with the enhancement factor
$4K_\eta$ for set II(f) discussed in Sec. \ref{sec:leading}.
%The logarithmic enhancement comes from the $v-p$ loop only.
%This is consistent with $r \simeq 11 - 15$ for (e,e), (me) and
%$r \simeq 0.7 - 2$ for (e,m), (m,m).\\
\label{table2fB}
}
\begin{tabular}{lcrrrr}
\hline
\hline
~Integral~~& ~~~$n_F$~~~ &~ ~Value (error)~~~&~Sampling~per~& ~~~No. of~~~\hspace{4mm} &~$r$~ \\ [.1cm]  
&  &~including~$n_F$~~~&~iteration~~~~~& ~~iterations~ \hspace{4mm} &  \\ [.1cm]   \hline
\\
$JKLp_2^{(e,e)}$&24&$-11.37774~(313)$&~~$1 \times 10^{7}, 1\times 10^9$ \hspace{4mm}~~&250, 20\hspace{14mm}~&12.35 \\
$JKLp_2^{(m,e)}$&24&$-1.87116~(~30)$&~~$1 \times 10^{7}$ \hspace{4mm}~~&250\hspace{14mm}~&15.86 \\
$JKLp_2^{(e,m)}$&24&$-0.74946~(~64)$&~~$1 \times 10^{7}$ \hspace{4mm}~~&250\hspace{14mm}~&0.81 \\
$JKLp_2^{(m,m)}$&24&$-0.28035~(~39)$&~~$1 \times 10^{7}$ \hspace{4mm}~~&250\hspace{14mm}~&2.38 \\
\\
$JKLp_{13}^{(e,e)}$&48&$-45.68188~(1079)$&~~$1 \times 10^{8},1\times 10^9$ \hspace{4mm}~~&450, 100\hspace{14mm}~&13.03 \\
$JKLp_{13}^{(m,e)}$&48&$-13.81420~(~386)$&~~$1 \times 10^{7}, 1\times 10^9 $ \hspace{4mm}~~&250, 40\hspace{14mm}~&15.82 \\
$JKLp_{13}^{(e,m)}$&48&$-3.96527~(~399)$&~~$1 \times 10^{7}$ \hspace{4mm}~~&250\hspace{14mm}~&1.13 \\
$JKLp_{13}^{(m,m)}$&48&$-2.15339~(~182)$&~~$1 \times 10^{7}$ \hspace{4mm}~~&250\hspace{14mm}~&2.47 \\
\\
\\
$A_2 [2f]^{(e,e)}$&72&-57.0596~(113)&~&~& \\
$A_2 [2f]^{(m,e)}$&72&-15.6854~(39)&~&~& \\
$A_2 [2f]^{(e,m)}$&72&-4.7147~(~40)&~&~& \\
$A_2 [2f]^{(m,m)}$&72&-2.4338~(~19)&~&~& \\
\\
\\
\hline
\hline
\end{tabular}
\end{table}
\renewcommand{\arraystretch}{1}

\renewcommand{\arraystretch}{1.0}
\begin{table}
\caption{Numerical evaluation of diagrams of Set VI(a)
and Set VI(b).   
The notation follows that of \cite{oldtk1} with some modification
and adaptation.  \\
\label{table6}
}
\begin{tabular}{lcrrr}
\hline
\hline
~Integral~~&~~~ $n_F$~~~ &~ ~Value (error)~~~&~Sampling~per~& ~~~No. of~~~ \\ [.1cm]  
&  &~including $n_F$~ ~&~iteration~~~~~& ~~iterations~  \\ [.1cm]   \hline
\\
$M_{6LL,p2p2}^{(e,e,e)}$&6&$542.91180~(910)$&~~$1 \times 10^{9}$ \hspace{4mm}~~&300\hspace{4mm}~ \\
$M_{6LL,p2p2}^{(e,e,m)}$&72&$~39.00349~(272)$&~~$1 \times 10^{8}$ \hspace{4mm}~~&200\hspace{4mm}~ \\
$M_{6LL,p2p2}^{(e,m,m)}$&36&$~2.43029~(~22)$&~~$1 \times 10^{8}$ \hspace{4mm}~~&200\hspace{4mm}~ \\
$M_{6LL,p2p2}^{(m,e,e)}$&36&$~34.42389~(680)$&~~$1 \times 10^{8}$ \hspace{4mm}~~&200\hspace{4mm}~ \\
$M_{6LL,p2p2}^{(m,e,m)}$&72&$~10.37125~(162)$&~~$1 \times 10^{8}$ \hspace{4mm}~~&200\hspace{4mm}~ \\
$M_{6LL,p2p2}^{(m,m,m)}$&36&$~1.04171~(~37)$&~~$1 \times 10^{8}$ \hspace{4mm}~~&20\hspace{4mm}~ \\
\\
$M_{6LL,p4}^{(e,e)}$&54&$168.72855~(478)$&~~$1 \times 10^{9}$ \hspace{4mm}~~&200\hspace{4mm}~ \\
$M_{6LL,p4}^{(e,m)}$&54&$~~7.58383~(~50)$&~~$1 \times 10^{8}$ \hspace{4mm}~~&210\hspace{4mm}~ \\
$M_{6LL,p4}^{(m,e)}$&54&$~~4.81614~(164)$&~~$1 \times 10^{8}$ \hspace{4mm}~~&200\hspace{4mm}~ \\
$M_{6LL,p4}^{(m,m)}$&54&$~~1.34726~(~30)$&~~$1 \times 10^{8}$ \hspace{4mm}~~&140\hspace{4mm}~ \\
\\
\hline
\hline
\end{tabular}
\end{table}

\renewcommand{\arraystretch}{0.8}
\begin{table}
\caption{Numerical evaluation of diagrams of Set VI(c).
The notation follows that of \cite{oldtk1} with some modification
and adaptation.  
The suffix $p$ indicates insertion of $\Pi_2$ in the photon lines
connecting the muon line and $l$-$l$ loop.
$r = \Delta M_{8LLxp}/\Delta M_{8LLx}$, where $x = E, F, G, H, I$,
in column 6 is for comparison with the crude enhancement factor
$3K_\eta$ for set VI(c) discussed in Sec. \ref{sec:leading}.\\
\label{table6c}
}
\begin{tabular}{lcrrrr}
\hline
\hline
~Integral~~& ~~~$n_F$~~~ &~ ~Value (error)~~~&~Sampling~per~& ~~~No. of~~~\hspace{4mm} &~$r$~ \\ [.1cm]  
&  &~including~$n_F$~~~&~iteration~~~~~& ~~iterations~ \hspace{4mm} &  \\ [.1cm]   \hline
\\
$\Delta M_{8LLEp}^{(e,e)}$&18&$-82.9940~(141)$&~~$2 \times 10^{8}$ \hspace{4mm}~~&200\hspace{14mm}~&3.84 \\
$\Delta M_{8LLEp}^{(e,m)}$&18&$-1.8277~(~18)$&~~$1 \times 10^{7}$ \hspace{4mm}~~&40\hspace{14mm}~& \\
$\Delta M_{8LLEp}^{(m,e)}$&18&$-3.3463~(~44)$&~~$1 \times 10^{7}$ \hspace{4mm}~~&140\hspace{14mm}~& \\
$\Delta M_{8LLEp}^{(m,m)}$&18&$-0.6168~(~~6)$&~~$1 \times 10^{7}$ \hspace{4mm}~~&140\hspace{14mm}~& \\
\\
$\Delta M_{8LLFp}^{(e,e)}$&36&$-322.4493~(~573)$&~~$2 \times 10^{8}$ \hspace{4mm}~~&280\hspace{14mm}~&4.25 \\
$\Delta M_{8LLFp}^{(e,m)}$&36&$-5.2571~(~60)$&~~$1 \times 10^{7}$ \hspace{4mm}~~&80\hspace{14mm}~& \\
$\Delta M_{8LLFp}^{(m,e)}$&36&$-9.3199~(~64)$&~~$4 \times 10^{7}$ \hspace{4mm}~~&390\hspace{14mm}~& \\
$\Delta M_{8LLFp}^{(m,m)}$&36&$-1.5392~(~17)$&~~$1 \times 10^{7}$ \hspace{4mm}~~&280\hspace{14mm}~& \\
\\
$\Delta M_{8LLGp}^{(e,e)}$&36&$-181.5345~(489)$&~~$2 \times 10^{8}$ \hspace{4mm}~~&210\hspace{14mm}~&5.17 \\
$\Delta M_{8LLGp}^{(e,m)}$&36&$-3.3841~(~61)$&~~$1 \times 10^{7}$ \hspace{4mm}~~&80\hspace{14mm}~& \\
$\Delta M_{8LLGp}^{(m,e)}$&36&$-6.5157~(~84)$&~~$4 \times 10^{7}$ \hspace{4mm}~~&300\hspace{14mm}~& \\
$\Delta M_{8LLGp}^{(m,m)}$&36&$-0.9499~(~18)$&~~$1 \times 10^{7}$ \hspace{4mm}~~&300\hspace{14mm}~& \\
\\
$\Delta M_{8LLHp}^{(e,e)}$&18&$230.3344~(590)$&~~$4 \times 10^{8}$ \hspace{4mm}~~&400\hspace{14mm}~&4.27 \\
$\Delta M_{8LLHp}^{(e,m)}$&18&$2.2093~(~74)$&~~$1 \times 10^{7}$ \hspace{4mm}~~&66\hspace{14mm}~& \\
$\Delta M_{8LLHp}^{(m,e)}$&18&$2.1519~(~96)$&~~$4 \times 10^{7}$ \hspace{4mm}~~&480\hspace{14mm}~& \\
$\Delta M_{8LLHp}^{(m,m)}$&18&$0.2839~(~18)$&~~$1 \times 10^{7}$ \hspace{4mm}~~&320\hspace{14mm}~& \\
\\
$\Delta M_{8LLIp}^{(e,e)}$&36&$514.7317~(567)$&~~$2 \times 10^{8}$ \hspace{4mm}~~&300\hspace{14mm}~&4.57 \\
$\Delta M_{8LLIp}^{(e,m)}$&36&$7.0544~(~59)$&~~$1 \times 10^{7}$ \hspace{4mm}~~&80\hspace{14mm}~& \\
$\Delta M_{8LLIp}^{(m,e)}$&36&$9.2489~(~68)$&~~$4 \times 10^{7}$ \hspace{4mm}~~&320\hspace{14mm}~& \\
$\Delta M_{8LLIp}^{(m,m)}$&36&$1.1912~(~13)$&~~$1 \times 10^{7}$ \hspace{4mm}~~&340\hspace{14mm}~& \\
\\
\hline
\hline
\end{tabular}
\end{table}
\renewcommand{\arraystretch}{1}

\renewcommand{\arraystretch}{0.8}
\begin{table}
\caption{Numerical evaluation of diagrams of Set VI(e).   
The notation follows that of \cite{oldtk1} with some modification
and adaptation.  
The suffix $q$ refers to
insertion of $\Pi_2$ in radiative corrections to the muon line.
$r = \Delta M_{8LLxp}/\Delta M_{8LLx}$, where $x = E, F, G, H, I$,
in column 6 is for comparison with the crude enhancement factor
 for set VI(e) discussed in Sec. \ref{sec:leading}.\\
\label{table6e}
}
\begin{tabular}{lcrrrr}
\hline
\hline
~Integral~~& ~~~$n_F$~~~ &~ ~Value (error)~~~&~Sampling~per~& ~~~No. of~~~\hspace{4mm} &~$r$~ \\ [.1cm]  
&  &~including~$n_F$~~~&~iteration~~~~~& ~~iterations~ \hspace{4mm} &  \\ [.1cm]   \hline
\\
$\Delta M_{8LLEq}^{(e,e)}$&6&$-35.1438~(~514)$&~~$1 \times 10^{7}$ \hspace{4mm}~~&80\hspace{14mm}~&1.62 \\
$\Delta M_{8LLEq}^{(e,m)}$&6&$0.0505~(16)$&~~$1 \times 10^{7}$ \hspace{4mm}~~&40\hspace{14mm}~& \\
$\Delta M_{8LLEq}^{(m,e)}$&6&$-0.9986~(17)$&~~$1 \times 10^{7}$ \hspace{4mm}~~&120\hspace{14mm}~& \\
$\Delta M_{8LLEq}^{(m,m)}$&6&$-0.1347~(2)$&~~$1 \times 10^{7}$ \hspace{4mm}~~&80\hspace{14mm}~& \\
\\
$\Delta M_{8LLFq}^{(e,e)}$&12&$-100.5201~(458)$&~~$4 \times 10^{7}$ \hspace{4mm}~~&220\hspace{14mm}~&1.33 \\
$\Delta M_{8LLFq}^{(e,m)}$&12&$0.3570~(19)$&~~$1 \times 10^{7}$ \hspace{4mm}~~&100\hspace{14mm}~& \\
$\Delta M_{8LLFq}^{(m,e)}$&12&$-2.4368~(41)$&~~$1 \times 10^{7}$ \hspace{4mm}~~&140\hspace{14mm}~& \\
$\Delta M_{8LLFq}^{(m,m)}$&12&$-0.2254~(3)$&~~$1 \times 10^{7}$ \hspace{4mm}~~&100\hspace{14mm}~& \\
\\
$\Delta M_{8LLGq}^{(e,e)}$&12&$-38.1520~(~399)$&~~$1 \times 10^{7}$ \hspace{4mm}~~&80\hspace{14mm}~&1.09 \\
$\Delta M_{8LLGq}^{(e,m)}$&12&$-0.2662~(~7)$&~~$1 \times 10^{7}$ \hspace{4mm}~~&40\hspace{14mm}~& \\
$\Delta M_{8LLGq}^{(m,e)}$&12&$-1.6144~(33)$&~~$1 \times 10^{7}$ \hspace{4mm}~~&180\hspace{14mm}~& \\
$\Delta M_{8LLGq}^{(m,m)}$&12&$-0.0957~(3)$&~~$1 \times 10^{7}$ \hspace{4mm}~~&80\hspace{14mm}~& \\
\\
$\Delta M_{8LLHq}^{(e,e)}$&6&$64.5209~(~879)$&~~$1 \times 10^{7}$ \hspace{4mm}~~&80\hspace{14mm}~&1.19 \\
$\Delta M_{8LLHq}^{(e,m)}$&6&$-0.4527~(12)$&~~$1 \times 10^{7}$ \hspace{4mm}~~&40\hspace{14mm}~& \\
$\Delta M_{8LLHq}^{(m,e)}$&6&$0.3954~(39)$&~~$1 \times 10^{7}$ \hspace{4mm}~~&160\hspace{14mm}~& \\
$\Delta M_{8LLHq}^{(m,m)}$&6&$-0.221~(91)\times 10^{-3}$&~~$1 \times 10^{7}$ \hspace{4mm}~~&80\hspace{14mm}~& \\
\\
$\Delta M_{8LLIq}^{(e,e)}$&12&$189.0518~(620)$&~~$4 \times 10^{7}$ \hspace{4mm}~~&200\hspace{14mm}~&1.68 \\
$\Delta M_{8LLIq}^{(e,m)}$&12&$1.8726~(~19)$&~~$1 \times 10^{7}$ \hspace{4mm}~~&80\hspace{14mm}~& \\
$\Delta M_{8LLIq}^{(m,e)}$&12&$2.0747~(~37)$&~~$1 \times 10^{7}$ \hspace{4mm}~~&200\hspace{14mm}~& \\
$\Delta M_{8LLIq}^{(m,m)}$&12&$0.0719~(~~2)$&~~$1 \times 10^{7}$ \hspace{4mm}~~&80\hspace{14mm}~& \\
\\
\hline
\hline
\end{tabular}
\end{table}
\renewcommand{\arraystretch}{1}

\renewcommand{\arraystretch}{1.0}
\begin{table}
\caption{Numerical evaluation of diagrams of Set VI(f).
The notation follows that of \cite{oldtk1} with some modification
and adaptation.  
The suffix $p$ indicates insertion of $\Pi_2$ in the photon lines
connecting the muon line and the $l-l$ loop.
$r = \Delta M_{8LLxp}/\Delta M_{8LLx}$, where $x = A, B, C, D$,
in column 6 is for comparison with the enhancement factor
$3K_\eta$ for set VI(f) discussed in Sec. \ref{sec:leading}.\\
\label{table6f}
}
\begin{tabular}{lcrrrr}
\hline
\hline
~Integral~~& ~~~$n_F$~~~ &~ ~Value (error)~~~&~Sampling~per~& ~~~No. of~~~\hspace{4mm} &~$r$~ \\ [.1cm]  
&  &~including~$n_F$~~~&~iteration~~~~~& ~~iterations~ \hspace{4mm} &  \\ [.1cm]   \hline
\\
$\Delta M_{8LLAp}^{(e,e)}$&30&$307.3206~(848)$&~~$4 \times 10^{7}$ \hspace{4mm}~~&200\hspace{14mm}~&5.90  \\
$\Delta M_{8LLAp}^{(e,m)}$&30&$8.9175~(~82)$&~~$1 \times 10^{7}$ \hspace{4mm}~~&120\hspace{14mm}~&  \\
$\Delta M_{8LLAp}^{(m,e)}$&30&$2.9097~(118)$&~~$1 \times 10^{7}$ \hspace{4mm}~~&100\hspace{14mm}~&  \\
$\Delta M_{8LLAp}^{(m,m)}$&30&$0.7878~(~10)$&~~$1 \times 10^{7}$ \hspace{4mm}~~&200\hspace{14mm}~&  \\
\\
$\Delta M_{8LLBp}^{(e,e)}$&60&$-482.5729~(603)$&~~$1 \times 10^{8}$ \hspace{4mm}~~&200\hspace{14mm}~&6.43 \\
$\Delta M_{8LLBp}^{(e,m)}$&60&$-16.0636~(~78)$&~~$1 \times 10^{7}$ \hspace{4mm}~~&120\hspace{14mm}~& \\
$\Delta M_{8LLBp}^{(m,e)}$&60&$-13.7242~(121)$&~~$2 \times 10^{7}$ \hspace{4mm}~~&100\hspace{14mm}~& \\
$\Delta M_{8LLBp}^{(m,m)}$&60&$-1.7449~(~12)$&~~$1 \times 10^{7}$ \hspace{4mm}~~&280\hspace{14mm}~& \\
\\
$\Delta M_{8LLCp}^{(e,e)}$&60&$645.3472~(823)$&~~$1 \times 10^{8}$ \hspace{4mm}~~&200\hspace{14mm}~&6.00 \\
$\Delta M_{8LLCp}^{(e,m)}$&60&$19.7833~(~82)$&~~$1 \times 10^{7}$ \hspace{4mm}~~&180\hspace{14mm}~& \\
$\Delta M_{8LLCp}^{(m,e)}$&60&$30.4954~(125)$&~~$2 \times 10^{7}$ \hspace{4mm}~~&240\hspace{14mm}~& \\
$\Delta M_{8LLCp}^{(m,m)}$&60&$3.4824~(~16)$&~~$1 \times 10^{7}$ \hspace{4mm}~~&300\hspace{14mm}~& \\
\\
$\Delta M_{8LLDp}^{(e,e)}$&30&$-252.4292~(616)$&~~$4 \times 10^{7}$ \hspace{4mm}~~&200\hspace{14mm}~&6.67 \\
$\Delta M_{8LLDp}^{(e,m)}$&30&$-8.4527~(~66)$&~~$1 \times 10^{7}$ \hspace{4mm}~~&80\hspace{14mm}~& \\
$\Delta M_{8LLDp}^{(m,e)}$&30&$-1.1736~(~86)$&~~$1 \times 10^{7}$ \hspace{4mm}~~&140\hspace{14mm}~& \\
$\Delta M_{8LLDp}^{(m,m)}$&30&$-0.4074~(~~8)$&~~$1 \times 10^{7}$ \hspace{4mm}~~&200\hspace{14mm}~& \\
\\
\hline
\hline
\end{tabular}
\end{table}
\renewcommand{\arraystretch}{1}

\renewcommand{\arraystretch}{1.0}
\begin{table}
\caption{Numerical evaluation of diagrams of Set VI(i).
The notation follows that of \cite{oldtk1} with some modification
and adaptation.  
The suffix $q$ refers to
insertion of $\Pi_2$ in radiative corrections to the $l-l$ loop.
$r = \Delta M_{8LLxp}/\Delta M_{8LLx}$, where $x = A, B, C, D$,
in column 6 is for comparison with the enhancement factor
$K_\eta$ for set VI(i) discussed in Sec. \ref{sec:leading}.\\
\label{table6i}
}
\begin{tabular}{lcrrrr}
\hline
\hline
~Integral~~& ~~~$n_F$~~~ &~ ~Value (error)~~~&~Sampling~per~& ~~~No. of~~~\hspace{4mm} &~$r$~ \\ [.1cm]  
&  &~including~$n_F$~~~&~iteration~~~~~& ~~iterations~ \hspace{4mm} &  \\ [.1cm]   \hline
\\
$\Delta M_{8LLAq}^{(e,e)}$&10&$29.6915~(199)$&~~$1 \times 10^{7}$ \hspace{4mm}~~&80\hspace{14mm}~&0.57 \\
$\Delta M_{8LLAq}^{(e,m)}$&10&$0.0914~(~3)$&~~$1 \times 10^{7}$ \hspace{4mm}~~&40\hspace{14mm}~& \\
$\Delta M_{8LLAq}^{(m,e)}$&10&$0.4867~(30)$&~~$1 \times 10^{7}$ \hspace{4mm}~~&100\hspace{14mm}~& \\
$\Delta M_{8LLAq}^{(m,m)}$&10&$0.8068~(93) \times 10^{-2}$&~~$1 \times 10^{7}$ \hspace{4mm}~~&140\hspace{14mm}~& \\
\\
$\Delta M_{8LLBq}^{(e,e)}$&20&$-81.3369~(946)$&~~$1 \times 10^{7}$ \hspace{4mm}~~&95*\hspace{14mm}~&1.08 \\
$\Delta M_{8LLBq}^{(e,m)}$&20&$-1.2816~(26)$&~~$1 \times 10^{7}$ \hspace{4mm}~~&100\hspace{14mm}~& \\
$\Delta M_{8LLBq}^{(m,e)}$&20&$-4.6043~(69)$&~~$1 \times 10^{7}$ \hspace{4mm}~~&138\hspace{14mm}~& \\
$\Delta M_{8LLBq}^{(m,m)}$&20&$-0.4926~(8)$&~~$1 \times 10^{7}$ \hspace{4mm}~~&200\hspace{14mm}~& \\
\\
$\Delta M_{8LLCq}^{(e,e)}$&20&$71.0664~(409)$&~~$1 \times 10^{7}$ \hspace{4mm}~~&80\hspace{14mm}~&0.66 \\
$\Delta M_{8LLCq}^{(e,m)}$&20&$0.7624~(~7)$&~~$1 \times 10^{7}$ \hspace{4mm}~~&40\hspace{14mm}~& \\
$\Delta M_{8LLCq}^{(m,e)}$&20&$8.5729~(59)$&~~$1 \times 10^{7}$ \hspace{4mm}~~&160\hspace{14mm}~& \\
$\Delta M_{8LLCq}^{(m,m)}$&20&$0.5664~(4)$&~~$1 \times 10^{7}$ \hspace{4mm}~~&100\hspace{14mm}~& \\
\\
$\Delta M_{8LLDq}^{(e,e)}$&10&$-43.7735~(449)$&~~$1 \times 10^{7}$ \hspace{4mm}~~&80\hspace{14mm}~&1.16 \\
$\Delta M_{8LLDq}^{(e,m)}$&10&$-0.6699~(14)$&~~$1 \times 10^{7}$ \hspace{4mm}~~&60\hspace{14mm}~& \\
$\Delta M_{8LLDq}^{(m,e)}$&10&$-0.2587~(42)$&~~$1 \times 10^{7}$ \hspace{4mm}~~&100\hspace{14mm}~& \\
$\Delta M_{8LLDq}^{(m,m)}$&10&$-0.0551~(5)$&~~$1 \times 10^{7}$ \hspace{4mm}~~&140\hspace{14mm}~& \\
\\
\hline
\hline
\end{tabular}
\end{table}
\renewcommand{\arraystretch}{1}

\renewcommand{\arraystretch}{1.0}
\begin{table}
\caption{Numerical evaluation of diagrams of Set VI(j)
contributing to the muon g-2.
In the superscript $(a,b)$, $a$ and $b$ refer to 
the external and internal light-by-light-scattering
loop, respectively.
Last 4 lines with superscript $(m,m)$ are 
mass-independent so that they are identical with the contributions
to the electron g-2.  \\
\label{table6j}
}
\begin{tabular}{lcrrr}
\hline
\hline
~Integral~~& ~~~$n_F$~~~&~Value (Error)~~~~~&~Sampling~per~& ~~~No. of~~~ \\ [.1cm] 
 & &~~including $n_F$~~~~~&~iteration~~~~~& ~~iterations~  \\ [.1cm]   \hline
\\
 $X6j_{a}^{(e,e)}$ & 24 ~~&    0.57928   (0.01386 ) &$ ~~~~1\times10^8$,$1\times10^9$ & 100, 294 \\
 $X6j_{b}^{(e,e)}$ & 12 ~~&  -16.91235   (0.00630 ) &$ ~~~~1\times10^8$,$1\times10^9$ & 100, 100 \\
 $X6j_{c}^{(e,e)}$ & 12 ~~&  -23.00801   (0.00777 ) &$~~~~ 1\times10^8$,$1\times10^9$ & 100, 160 \\
 $X6j_{d}^{(e,e)}$ &  6 ~~&   15.38181   (0.00388 ) &$~~~~ 1\times10^8$,$1\times10^9$ & 100,  87 \\
\\
 $X6j_{a}^{(e,m)}$ & 24 ~~&    4.75211   (0.00609 ) &$ 1\times10^8$    & 105 \\
 $X6j_{b}^{(e,m)}$ & 12 ~~&   -0.84570   (0.00112 ) &$ 1\times10^8$    & 100 \\
 $X6j_{c}^{(e,m)}$ & 12 ~~&   -5.96339   (0.00157 ) &$ 1\times10^8$    & 190 \\
 $X6j_{d}^{(e,m)}$ &  6 ~~&    0.88153   (0.00046 ) &$ 1\times10^8$    & 220 \\
\\
 $X6j_{a}^{(m,e)}$ & 24 ~~&   -2.06921   (0.00549 ) &$ 1\times10^8$    & 100 \\
 $X6j_{b}^{(m,e)}$ & 12 ~~&   -3.75200   (0.00232 ) &$ 1\times10^8$    & 100 \\
 $X6j_{c}^{(m,e)}$ & 12 ~~&    1.64453   (0.00162 ) &$ 1\times10^8$    & 200 \\
 $X6j_{d}^{(m,e)}$ &  6 ~~&    3.80656   (0.00138 ) &$ 1\times10^8$    & 130 \\
\\
 $X6j_{a}^{(m,m)}$         & 24 ~~&   -0.22601   (0.00143 ) &$ 1\times10^8$      &  180 \\ 
 $X6j_{b}^{(m,m)}$         & 12 ~~&   -0.69698   (0.00065 ) &$ 1\times10^8$      &  100 \\
 $X6j_{c}^{(m,m)}$         & 12 ~~&   -0.02753   (0.00053 ) &$ 1\times10^8$      &  170 \\
 $X6j_{d}^{(m,m)}$         &  6 ~~&    0.72170   (0.00037 ) &$ 1\times10^8$      &  100 \\
\\
\hline 
\hline
\end{tabular}
\end{table}
\renewcommand{\arraystretch}{1}

\renewcommand{\arraystretch}{1.0}
\begin{table}
\caption{Numerical evaluation of diagrams of Set VI(k)
contributing to the muon g-2.
The superscript $e$ ($m$) indicates that the lepton loop $\Lambda_6$
is the electron (muon) loop.
The latter is mass-independent so that it is identical with the contribution
to the electron g-2.  \\
\label{table6k}
}
\begin{tabular}{lcrrr}
\hline
\hline
~Integral~~& ~~~$n_F$~~~&~Value (Error)~~~~~&~Sampling~per~& ~~~No. of~~~ \\ [.1cm] 
 & &~~including $n_F$~~~~~&~iteration~~~~~& ~~iterations~  \\ [.1cm]   \hline
\\
 $X6k_{a}^{(e)}$& 10~~ &   50.35921   (0.01998 ) & $1\times10^8, 1\times10^9 $   
& 500 , 150\\
 $X6k_{b}^{(,e)}$& 10~~ &    1.72669   (0.01786 ) & $1\times10^8 $     & 400 \\
 $X6k_{c}^{(e)}$& 20~~ &    7.81330   (0.02038 ) & $1\times10^8 $     & 300 \\
 $X6k_{d}^{(e)}$& 20~~ &   20.67840   (0.03758 ) & $1\times10^8 $     & 100 \\
 $X6k_{e}^{(e)}$& 10~~ &   -0.19466   (0.01045 ) & $1\times10^8 $    & 300 \\
 $X6k_{f}^{(e)}$& 10~~ &    1.75890   (0.02374 ) & $1\times10^8 $    & 230 \\
 $X6k_{g}^{(e)}$& 20~~ &   -0.02607   (0.01797 ) & $1\times10^8 $    & 200 \\
 $X6k_{h}^{(e)}$& 10~~ &   -0.69054   (0.00750 ) & $1\times10^8 $    & 100 \\
 $X6k_{i}^{(e)}$& 10~~ &   15.69736   (0.01595 ) & $1\times10^8 $    & 200 \\
\\
 $X6k_{a}^{(m)}$& 10~~ &   -0.56022   (0.00301 ) & $1\times10^8 $   &  100 \\ 
 $X6k_{b}^{(m)}$& 10~~ &    0.30282   (0.00085 ) & $1\times10^8 $     &  100 \\
 $X6k_{c}^{(m)}$& 20~~ &   -0.32547   (0.00114 ) & $1\times10^8 $     &  100 \\
 $X6k_{d}^{(m)}$& 20~~ &    0.82380   (0.00084 ) & $1\times10^8 $     &  100 \\
 $X6k_{e}^{(m)}$& 10~~ &   -0.17188   (0.00053 ) & $1\times10^8 $     &  100 \\ 
 $X6k_{f}^{(m)}$& 10~~ &    0.30329   (0.00088 ) & $1\times10^8 $     &  100 \\
 $X6k_{g}^{(m)}$& 20~~ &   -0.94843   (0.00067 ) & $1\times10^8 $     &  100 \\
 $X6k_{h}^{(m)}$& 10~~ &   -0.13877   (0.00018 ) & $1\times10^8 $     &  100 \\
 $X6k_{i}^{(m)}$& 10~~ &    1.39510   (0.00069 ) & $1\times10^8 $     &  100 \\
\\
\hline 
\hline
\end{tabular}
\end{table}
\renewcommand{\arraystretch}{1}

\end{document}